# Influence of Economic Decoupling in assessing carbon budget quotas for the European Union

Ilaria Perissi[a]* and Aled Jones[a]

[a] *Global Sustainability Institute, Anglia Ruskin University, Cambridge CB1 1PT, UK*

Email: ilaria.perissi1@aru.ac.uk*; aled.jones@aru.ac.uk

In the present study, for the first time, an effort-sharing approach based on Inertia and Capability principles is proposed to assess European Union (EU27) carbon budget distribution among the Member States. This is done within the context of achieving the Green Deal objective and EU27 carbon neutrality by 2050. An in-depth analysis is carried out about the role of Economic Decoupling embedded in the Capability principle to evaluate the correlation between the expected increase of economic production and the level of carbon intensity in the Member States. As decarbonization is a dynamic process, the study proposes a simple mathematical model as a policy tool to assess and redistribute Member States' carbon budgets as frequently as necessary to encourage progress or overcome the difficulties each Member State may face during the decarbonization pathways.

## Introduction

The problem of sharing the global mitigation effort to tackle climate change among countries is extensively addressed in the literature, from the perspectives of equity, international policy, economics and financing [1], [2], [3], [4], [5]. To contain global warming within 2°C above pre-industrial levels, as committed by the Paris Agreement [6], the remaining carbon budget (CB) [7] needs to be shared among participants in a social-economic-ecological system by distributing it according to a set of observable metrics.

There are several principles on which carbon budget partition can be made among the countries: based on the historical trends or "Inertia", which reflects a business as usual emissions scenarios [8]; based on demographic consistency, in which the quotas are proportioned according to the countries' population, also called "Equity" [9]; or based on economic wealth, also called "Capability" [10], which considers the Gross Domestic Product (GDP) as an index of the potential ability to pay for mitigation. Usually, any of those principles taken individually is sufficient to elaborate a reliable 'effort sharing' approach which is represented by a blend of these main drivers, resulting in an effective model able to explore future emission scenarios and assess the achievability of decarbonization goals.

In the present study, for the first time, an effort sharing approach based on Inertia and Capability principles is used to assess European Union (EU27) carbon budget distribution among the Member States to accomplish the Green Deal objective and EU27 carbon neutrality by 2050. The Equity principle, which suggests a democratic sharing based on country population, is not directly involved in the portioning but is considered limited in the Capability term, shaped using GDP per capita instead of total GDP of the country.

An in-depth analysis is carried out on the role of Economic Decoupling [11] which represents a further novelty the authors embedded in the Capability principle. The GDP trend, taken as a measure of the economic potential to invest in mitigation, should be placed in relationship with the greenhouse gas (GHG) emission rate trend to evaluate the correlation between the economic production and the level of carbonization still present in Member States' economies.

The study results in a simple mathematical model as an effort sharing based policy tool able to assess and redistribute Member States' carbon budgets as frequently as necessary to encourage decarbonization progress or to overcome the difficulties each Member State may face during the decarbonization pathways. At present, the model shows that the Inertia of higher and lower emitter Member States is still the predominant factor that affects the whole distribution of the EU27 carbon budget among the Member States, with only minor effects due to the countries' economic capability and economic decoupling.

**Methods**

The methods section is organized into different subsections: information about the data resources used to develop the study; the methodology to assess the emissions trajectories, necessary to evaluate the carbon budgets for EU27 as a whole, for Effort Sharing Regulation (ESR) and Emissions Trading System (ETS); methodologies to define Inertia, Capability and Decoupling terms that shape the effort sharing model, used to distribute the carbon budget between the Member States. Once defined, the effort sharing model will be used to assess and compare Member States' carbon budget allocations in two different studies: firstly, "Study 2016-2019" which concerns Inertia, Capability and Decoupling obtained by averaged data along 2016-2019; and secondly, "Study 2019", which concerns Inertia, Capability and Decoupling obtained by the most recent 2019 data. The comparative analysis between the two Studies aims to assess the evolution in the Members States Carbon Budgets (CBs) distribution from an averaged historical post-Paris Agreement (2016-2019) to the most recent available data (2019) that maps the Members States CBs at the start of the time associated with Green Deal decarbonization objectives.

*Data Resources*

Statistical data, academic literature and European Commission documents are the main sources used in this research. In particular:

- Statistics data are from Eurostat statistics [12] which is the most extensive database for Europe, including GHG emission and GDP data that are fundamental to building the blended mathematical model
- EC commission documents like Regulation [13], Climate Law [14] and ETS [15] are used as sources for ETS and ESR emissions allocations and budgets.

All sources are referred to in the text when appropriate.

*Estimation of European Union GHG emissions projections from 2020 to 2050*

Current GHG emissions for EU27, ESR and ETS sectors are used as a starting point which is then extrapolated to 2030 and 2050 to be in line with the Green Deal target of -55% GHG emissions rate of 1990 emissions in 2030 and the long-term target of a carbon neutral European Union by 2050. Once these projections are set it is possible to assess the carbon budget for EU27 as a whole and ESR and the ETS sectors. The extrapolations are in Figure 2. The next paragraphs detail how those trajectories are obtained.

*EU 27 total GHG projections*

A first linear interpolation of the EU27 GHG historical trends 1990-2018, available at the time of writing [16] has been performed to estimate emissions rates in 2019 and 2020 on the way to end the total EU27 GHG series in the same year (2020) as the ESR and ETS historical series. Note that 2020 total GHG emissions are overestimated due to the impact of lockdown measures used to tackle COVID19, with emissions expected to be less than this linear approximation. However, this decrease will be compensated by an emissions rebound [17] at the end of the pandemic emergency. This uncertainty on emissions historical trends now does not affect the reliability of the estimation of the cumulative Carbon Budget remaining to 2050.

Projections for EU27 GHG to 2050 are obtained by a second linear extrapolation from the year 2020 to the year 2050 and assumed to reach zero emissions. The emissions for the EU27 are found to be 3875,48 Mt $CO_2eq$ in 2020 from the first linearization. This results in an average of 129 Mt $CO_2eq$/year of emission decrease required between 2020 and 2050.

*EU27 ESR projections*

Historical GHG emissions data for ESR sectors are from Eurostat [18] in Table A.1 (Appendix A). Projections of GHG emissions to 2030 were initially taken from the Regulation [19] that set the target for EU27 ESR as -30% compared to 2005 emissions. The Green Deal's new objectives resulted in the Regulation being amended in July 2021 [20] with a EU27 Green Deal objective for ESR as -40% of 2005 emissions values in 2030 (instead of -30%). The amendment reports the new percentage of the final target for each Member State, but it does not report the annual allocations from 2021 to 2030 (as in ANNEX II of Regulation, see Table A.2 in Appendix A). However, the annual emissions allocation between 2021 and 2030 for each Member State within the Green Deal targets (-40% of 2005 in 2030) can be easily proportioned (see Table A.3, Appendix A) using the initial calculations with the original Regulation objectives (-30% of 2005 in 2030) and considering the final percentage ratio between emissions in 2030 for Green Deal and the emission at year 2030 for Regulation. Linear extrapolation from the last data available for ESR in 2030 (1513 Mt $CO_2eq$) to 2050 results in a 76 Mt $CO_2eq$/year decrease in GHG between 2030 and 2050 for ESR.

*EU27 ETS emissions projections*

Historical data for the ETS sector are from European Energy Agency [15]. The impact of ETS (revised in 2018) in force before the Green Deal was estimated [21] as delivering a 43% reduction in emissions for those sectors it covers by 2030 compared to 2005, consistent with a EU27 emissions reduction target of at least 40 % by 2030 [22] compared to 1990. To align the EU ETS Directive with the increased GHG emission reduction targets set in the European Climate Law, the Commission is proposing to reduce the emissions from the EU ETS sectors by 61% by 2030, compared to 2005 levels [21], an 18 percentage point increase from the previous target. To achieve this new target, the proposal increases the linear emissions reduction factor from 2.2% per year to 4.2%. From this study, the estimation of ETS future projection is linearly extrapolated to 1224 MtCO2eq in 2020 resulting in a 41 MtCO2eq/year GHG reduction between 2020 and 2050 to achieve a zero-emission target.

*Evaluation of Member States' Capability to invest in mitigation.*

Once a Carbon Budget for the EU27 has been assessed, the next step is to allocate quotas among the Member States. Here we rank the Member States on their capability to invest for mitigation, based on the "ability to pay" [23] [24] concept, used to distribute taxation across individuals with different incomes. Here the concept is used in a broader context, not related to taxation but using a more general definition of "capability to invest" in mitigation, or simply "capability" according to the IPCC's broad classification statement: "the greater an agent's ability to pay for the solution of a problem (here the problem of reaching the 2030 EU reduction goal) the greater the proportion of the costs that the agent should be expected to pay". Typically, the capability can be expressed by indicators based on GDP or GDP capita. Here we refer to the Capability Index (CI) proposed by Steininger et al. [25]:

$$CI_{MSj} = \frac{\frac{GDP_{capEU}}{GDP_{capMSj}}}{\sum_{1}^{j}\frac{GDP_{capEU}}{GDP_{MSi}}} \quad (1)$$

The ratio in Eq. 1 describes the deviation of GDP per capita of a Member State ($GDP_{capMS}$) from the EU-27 average GPD per capita ($GDP_{capitaEU}$); thus, a country carbon budget quota according to Capability is (Eq.2):

$$CB_{MSjcap} = CB_{EU27}\, CI_{MSj} = CB_{EU27}\, \frac{\frac{GDP_{capEU}}{GDP_{capMSj}}}{\sum_{1}^{j}\frac{GDP_{capEU}}{GDP_{MSi}}} \quad (2)$$

where $CB_{MSj}$ is the carbon budget of the Member State j, with j=27; $CB_{EU27}$ is the Carbon Budget for EU27 as a whole. GDP per capita are from Eurostat [26]. $CI_{MSj}$ assumes that higher GDP per capita countries can help the countries with a lower GDP per capita, but it does not consider other benefits/incentives that should be considered as well. However, GDP per capita is still the most widely used indicator in the literature and, more specifically, the EU Commission in allocating ESR quotas as a proxy to classify the various countries' economic capacities, at least for a first evaluation. $CI_{MSj}$ will be calculated with 2016-2019 average data for GDP capita (EU and countries) in Study 2016-2019 and with 2019 data in Study 2019.

*Evaluation of Economic Decoupling of the Member States*

GDP is one of the most used indexes to evaluate the wealth of a country and then its economic capability. However, in a decarbonization pathway, it is necessary to investigate if the GDP trend is still bound, and with which degree of intensity, to emission productions. The relationship between GHG and GDP is called carbon intensity and any deviation away from the correlated growth trajectory is an "Economic Decoupling" [27] [28].

There are several indexes used to measure decoupling; one of the simplest and most used is the Tapio [29] defined as (Eq.3):

$$\text{Tapio index} = \frac{\Delta\ \text{GHG}\%}{\Delta\ \text{GDP}\ \%} \qquad (3)$$

At the time of writing data for EU27 GDP capita are available until 2021 while data on GHG per capita stopped in 2019, the Tapio index for this research was calculated up to 2019. Moreover, considering the Paris Agreement commitment was signed in December 2015, the decoupling index was calculated back to 2016. Three historical ranges were considered for GDP and GHG: from 2016 to 2019 3 years), 2017-2019 (2 years) and 2018-2019 (1 year). Multiple evaluations of the Tapio index are necessary because a calculation of a Tapio index for each interval of time accounts for fluctuations that may occur in either GDP or GHG in each year which could artificially influence the comparison between specific years. Therefore, the final Tapio index is calculated as the average of the 3,2-, and 1-year indexes.

Notoriously, the Tapio index can assume positive and negative values, depending on decreasing GDP or GHG in the numerator and/or denominator of Eq. 3 and it is not suitable to be used as it is to create a portioning index to split carbon budgets. A positive Tapio index implies both GDP and growth over the interval (or both shrink). A negative Tapio index implies a stronger decoupling as GHG emissions shrink while GDP grows. To build a suitable portioning index, the decoupling factors for the EU27 countries need to be all positive or all negative. Thus, the authors rescaled the Tapio index for the lower decoupling index found among the Member States (that is the one from Malta). The Malta Tapio index in Study 2016-2019 is equal to 0.25 and 0.46 in Study 2019 which was rescaled to obtain a new index for Malta equal to -1 (see example in Figure 1- 1A and 1B). All other countries were then appropriately rescaled in the same way.

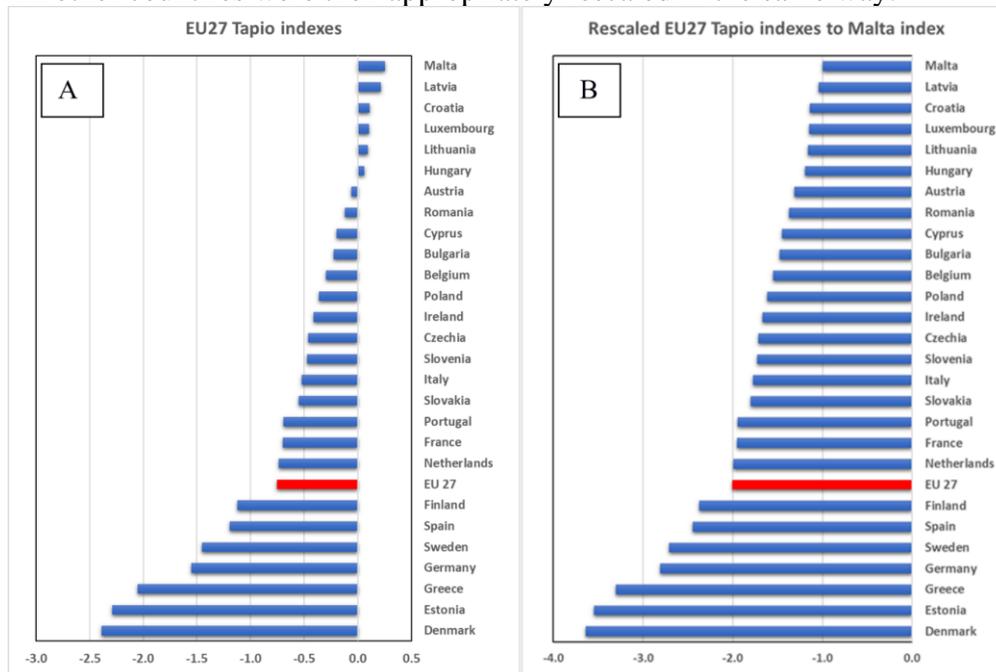

**Figure 1.** Example of rescaling Tapio index in the case of Study 2016-2019. (A) Averaged Tapio Index for the EU27 Member States; (B) Averaged Tapio index rescaled with respect to the averaged Malta Index equal to 0.25. Similar trends are obtained for Study 2019, where the Malta index is equal to 0.46.

Adopting the previous methodology, all Member States' rescaled Tapio indexes are negative and suitable to be used in creating the Decoupling index ($DI_{MSj}$) with the following ratio (Eq. 4):

$$DI_{MSj} = \frac{DI_{MSj}}{DI_{EU27}} \quad (4)$$

Where $DI_{MSj}$ are the rescaled Tapio index for the Member States j and $DI_{EU27}$ is the rescaled Tapio index for EU27. DI can now be used to build the proportioning "decoupling" index as (Eq. 5):

$$CB_{MSjdec} = CB_{EU27}\, DI_{MSj} = CB_{EU27}\, \frac{\frac{DI_{EU27}}{DI_{msj}}}{\sum_1^j \frac{1 DI_{EU27}}{DI_{msj}}} \quad (5)$$

*Evaluation of Inertia of the Member States*
Based on the principle of portioning the carbon budget according to past emissions, an Inertia Index ($II_{MSj}$) was calculated as the ratio between emissions of a Member States divided by the emission of EU27 (considering 2016-2019 data for the averaged study and data in 2019 for the most recent study). The carbon budget quotas due to the past emissions of the country are then obtained as follows (Eq. 6):

$$II_{MSj} = \frac{GHG_{MSj}}{GHG_{EU27}} \quad (6)$$

Then the Carbon budget quotas for each member states due to the inertia term is (Eq. 7):

$$CB_{MSjine} = CB_{EU27}\, II_{MSj} = CB_{EU27}\, \frac{GHG_{MSj}}{GHG_{EU27}} \quad (7)$$

*Carbon Budget blending approach*
The blending of Inertia, Capability and Decoupling contributions is based on the approach of Rapuach et al. [30]: employing the use of variable parameters, here indicated by w and z that span from 0 to 1 (with a 0.1 step) the blend is expressed as in the following equations. Eq. 8 considers the first blend with only Inertia and Capability:

$$CB_{MSjblend1} = (1-w)\, CB_{MSjcap} + w\, CB_{MSjine} \quad (8)$$

And, also considering the effect of decoupling, the second blend becomes as in Eq. 9:

$$CB_{MSjblend2} = (1-z)\, (w\, CB_{MSjdec} + (1-w)\, CB_{MSjcap}) + (z\, CB_{MSjine}) \quad (9)$$

Carbon Budgets blends are solved with Matlab ® using the previous equations in the vectorial form.

**Results and Discussion.**
The results and discussion section is organized into 6 subsections: the first section reports the results of the assessment of EU27 carbon budgets (as a whole, ESR and ETS), then an analysis of how the EU27 as a whole budget can be split among the countries based on of Capability, Decoupling and their blend; adding Inertia to the Capability Decoupling blend (Economic Term) and analysis of the blending results in Study 2016-2019 and Study 2019; Carbon Budgets estimation for the EU27 Member States.

*EU27 yearly emissions allocation and carbon budgets*

It is now possible to assess the carbon budgets of EU27, for ESR and ETS sectors from 2020 to 2030 and 2050 (Figure 2). Data are reported in Table 1.

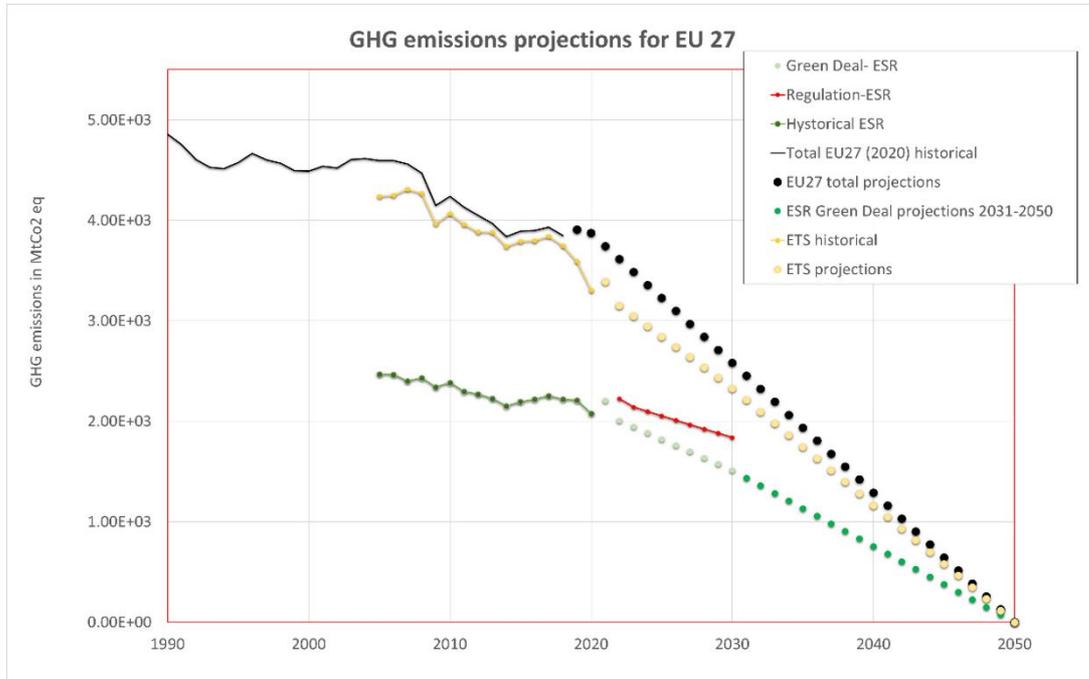

**Figure 2.** Linear extrapolation for GHG emissions for EU27, ESR and EST sectors, assessed in this study to accomplish Green Deal objectives.

The EU27 total emissions budget is the sum of the yearly allocated quotas by linear extrapolation. The carbon budget for the ESR sector to 2030 is evaluated according to the Green Deal target, this is the sum of the allocations reported in Table A.4 at the line "total" for EU27. Projections to 2050 are calculated as a linear extrapolation between allocation in 2030 (provided in Table A.4, considering -55% Green Deal goal) minus allocation in the year 2050 (that corresponds to 0). Similarly, the carbon Budget for ETS is a linear extrapolation between the allocation in the years 2020 and 2050.

The EU27 GHG yearly allocation is compatible with the study from Wolf et al. [31] which considers a more stringent decline to 2030 (162 MtCO2 eq/year instead of 129 MtCO2 from the present study) as an effect of the Green Deal Investment streams for 2021-2030, but a gentler decline between 2030 and 2050 (97 MtCO2eq) which was assessed with no specific measures from the Long-Term [32]. The estimated total carbon budget for EU27 from 2020 to 2050 is around 60 GtCO2 eq, which is compatible with the 2°C scenario (2015-2050) by Duscha et al. [33] and with the 2°C scenario (2011-2050) by Perissi et al. [34]. the authors are aware that those budgets for EU suffer from limitations and uncertainties due to considering mainly anthropogenic emissions factors, as recently raised in the study of Matthews et al [35] and Allen et al.[36]. 60 GtCO2eq does not pretend of being accurate enough as the definitive carbon budget goal for EU27 but this is functional to illustrate the authors' model for carbon budget portioning.

As a further observation, Regulation objectives (red line) were too weak to allow for complete decarbonization of the EU by 2050.

**Table 1** GHG projections and Carbon Budgets for ESR, ETS and EU total emissions to 2050 (Mt CO2 eq).

| year | ESR - yearly allocation | Total GHG EU27 | ETS |
|---|---|---|---|
| 2020 | 2079.17 | 3875.48 | 1224.24 |
| 2021 | 2206.66 | 3746.30 | 1183.43 |
| 2022 | 2008.38 | 3617.11 | 1142.62 |
| 2023 | 1946.90 | 3487.93 | 1101.82 |
| 2024 | 1885.42 | 3358.75 | 1061.01 |
| 2025 | 1823.95 | 3229.57 | 1020.20 |
| 2026 | 1762.47 | 3100.38 | 979.39 |
| 2027 | 1700.99 | 2971.20 | 938.58 |
| 2028 | 1639.51 | 2842.02 | 897.77 |
| 2029 | 1578.03 | 2712.84 | 856.97 |
| 2030 | 1512.66 | 2583.65 | 816.16 |
| 2031 | 1437.03 | 2454.47 | 775.35 |
| 2032 | 1361.40 | 2325.29 | 734.54 |
| 2033 | 1285.76 | 2196.11 | 693.73 |
| 2034 | 1210.13 | 2066.92 | 652.93 |
| 2035 | 1134.50 | 1937.74 | 612.12 |
| 2036 | 1058.86 | 1808.56 | 571.31 |
| 2037 | 983.23 | 1679.37 | 530.50 |
| 2038 | 907.60 | 1550.19 | 489.69 |
| 2039 | 831.97 | 1421.01 | 448.88 |
| 2040 | 756.33 | 1291.83 | 408.08 |
| 2041 | 680.70 | 1162.64 | 367.27 |
| 2042 | 605.07 | 1033.46 | 326.46 |
| 2043 | 529.43 | 904.28 | 285.65 |
| 2044 | 453.80 | 775.10 | 244.84 |
| 2045 | 378.17 | 645.91 | 204.04 |
| 2046 | 302.53 | 516.73 | 163.23 |
| 2047 | 226.90 | 387.55 | 122.42 |
| 2048 | 151.27 | 258.37 | 81.61 |
| 2049 | 75.63 | 129.18 | 40.80 |
| 2050 | 0.00 | 0.00 | 0.00 |
| **Carbon budget** | **34514.46** | **60069.94** | **18975.80** |

Notably, the summation of ESR and ETS budgets do not match exactly the cumulative one from EU27 historical emissions. This may be due to different data sources: compare for instance EU27 GHG emission from UNFCCC (UNFCCC, 2019) or Our World in Data [37] sources, which show a very similar trend but with slight differences in the absolute values. Based on the latest IPCC [38] reports, an almost linear relationship exists between cumulative CO2 emissions and increases in temperature, a relationship which, however, contains a probabilistic part that calculates a global carbon budget to achieve the 2°C global warming limit with a probability of 66%. This probabilistic component relates to uncertainty in climate sensitivity and therefore assessing an allowed carbon budget accurately is not currently possible. Therefore, for this study, we note that the difference shown by cumulative data for the ETS and ESR budgets and the Total EU27 GHG is around 10% (Table 1), an uncertainty below the overall uncertainty in the carbon budget as calculated by the IPCC methods, and therefore the authors choose to use the Eurostat sources, as all statistics then come from the same provider, and the differences between providers are assumed to be less relevant to the overall trends and narrative findings.

*Member States ranking according to their Economic Capability.*

Based on the Effort Sharing approach explained in the method section (Eq. 2), the allocation of Member States' carbon budgets because of their economic capability is reported in Table A.8. Results of Carbon Budget allocation for each Member States according to the Capability criterion are shown in Figure 3.

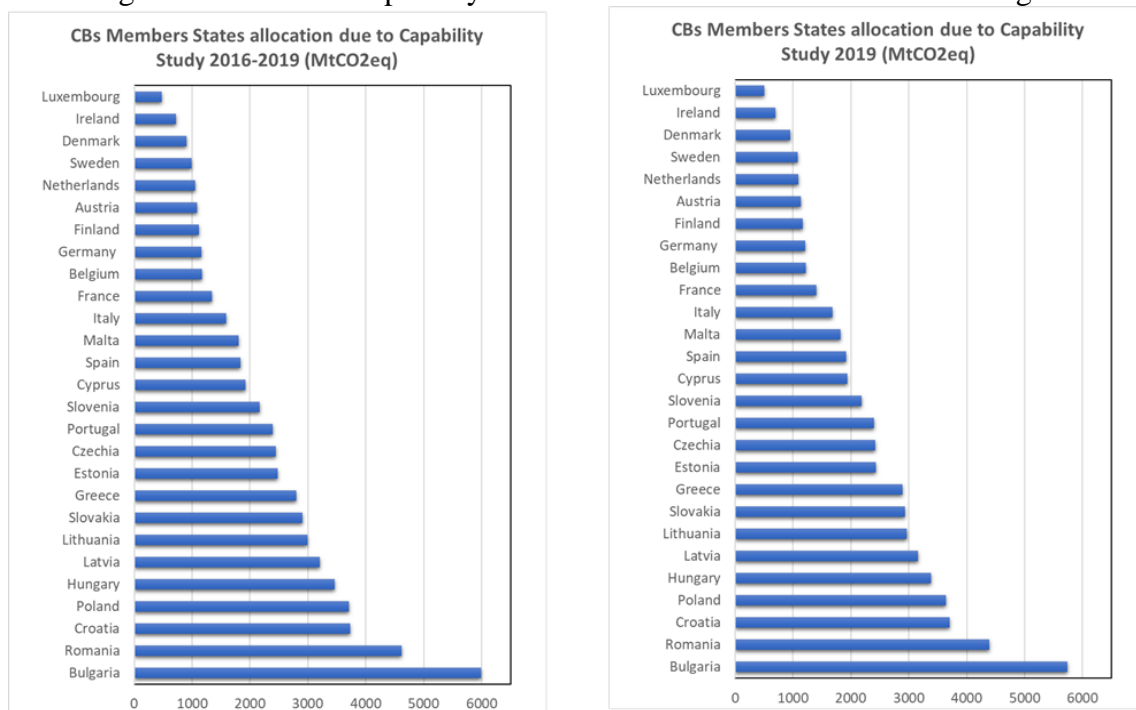

**Figure 3**. Member States CBs allocation according to the economic capability principle evaluated within 2016-2019 (left) and Study 2019 (right).

The ranking reflects the allocation of the EU 27 carbon budget among the Member States only based on their GDP per capita: the larger the carbon budget the smaller the ability to pay. As it will be shown in the next paragraphs, this allocation, as it is, has very limited applications, except for countries that have very similar inertia and economic decoupling terms, which is not the case for EU27.

*Decoupling of EU27 Member States Economies*

The Decoupling factors according to the Tapio index in Eq.2 have been calculated between the years 2016-2019, years 2017-2019, and years 2018-2019 (see Table A.7) and the related Tapio index are reported in Table 2

**Table 2.** Tapio Index for the EU27 countries, assessed for 3 years (2016 to 2019), 2 years (2017-2019) and 1 year (2018-2019) and their average.

| Countries | Tapio Index 2016-2019 | Tapio Index 2017-2019 | Tapio Index 2018-2019 | Member States Tapio index (average) |
|---|---|---|---|---|
| **European Union - 27** | -0.41 | -0.83 | -1.02 | -0.75 |
| **Belgium** | -0.19 | -0.15 | -0.55 | -0.30 |
| **Bulgaria** | -0.08 | -0.37 | -0.24 | -0.23 |
| **Czechia** | -0.22 | -0.42 | -0.75 | -0.46 |
| **Denmark** | -1.61 | -2.06 | -3.49 | -2.39 |
| **Germany** | -1.16 | -1.57 | -1.93 | -1.55 |
| **Estonia** | -0.99 | -1.96 | -3.92 | -2.29 |
| **Ireland** | -0.24 | -0.35 | -0.66 | -0.42 |
| **Greece** | -0.82 | -2.24 | -3.10 | -2.05 |
| **Spain** | -0.40 | -1.13 | -2.05 | -1.19 |
| **France** | -0.54 | -1.04 | -0.52 | -0.70 |
| **Croatia** | 0.17 | -0.13 | 0.28 | 0.11 |
| **Italy** | -0.41 | -0.35 | -0.81 | -0.52 |
| **Cyprus** | -0.10 | -0.32 | -0.18 | -0.20 |
| **Latvia** | 0.30 | 0.34 | 0.00 | 0.21 |
| **Lithuania** | 0.10 | 0.00 | 0.18 | 0.09 |
| **Luxembourg** | 0.31 | 0.28 | -0.28 | 0.10 |
| **Hungary** | 0.18 | 0.00 | 0.00 | 0.06 |
| **Malta** | 0.30 | 0.00 | **0.46** | **0.25** |
| **Netherlands** | -0.65 | -0.76 | -0.80 | -0.74 |
| **Austria** | -0.11 | -0.47 | 0.41 | -0.06 |
| **Poland** | -0.08 | -0.38 | -0.63 | -0.36 |
| **Portugal** | -0.10 | -1.00 | -0.97 | -0.69 |
| **Romania** | 0.05 | -0.08 | -0.34 | -0.12 |
| **Slovenia** | -0.25 | -0.41 | -0.76 | -0.48 |
| **Slovakia** | -0.17 | -0.46 | -1.01 | -0.55 |
| **Finland** | -0.75 | -0.50 | -2.12 | -1.12 |
| **Sweden** | 6.87 | 1.94 | -13.18 | -1.46 |

Tapio's interpretation of decoupling is reported in Tapio's original [29] and other papers [39], [11]. As a general rule "a strong decoupling" occurs when, ΔGDP>0 and ΔGHG <0, GDP increases and GHGs decrease and the Tapio index is negative; a "weak decoupling" when ΔGDP > 0, ΔGHG > 0 and ΔGDP > ΔGHG, GDP growth is faster than GHG, but their ratio should be no higher than 0.8 according to decoupling framework by Tapio. All EU27 Member States belong to one of these two series, thus we will not discuss here the other values the Tapio index might assume (positive but higher than 0.8, or GDP and GHG both negatives, for instance).

Looking at Table 2, the number of countries with a weak decoupling over three years are 8, in two years are 6, in the 1 year still 6. Decoupling graphical representations for each Member State are reported in the following Figure 4. GHG and GDP are set as 100 for the year 2010 and all the successive values are rescaled to this reference (see Table A.5 and Table A.6).

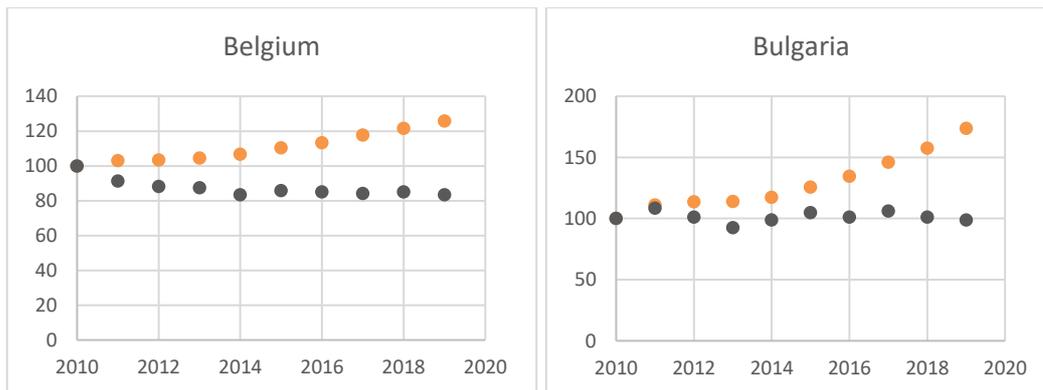

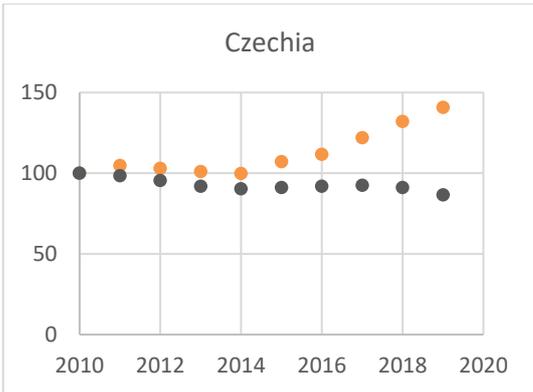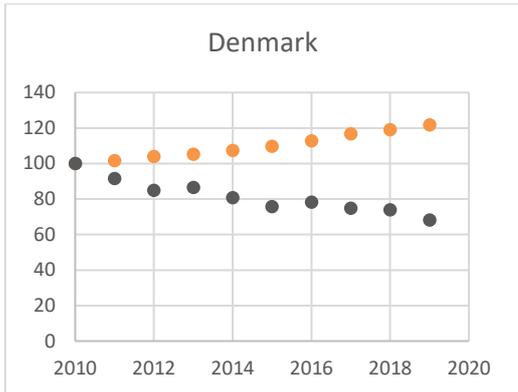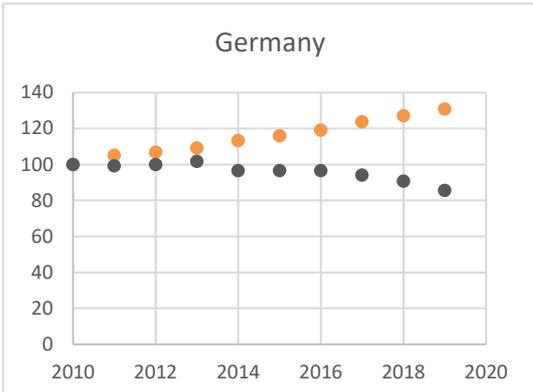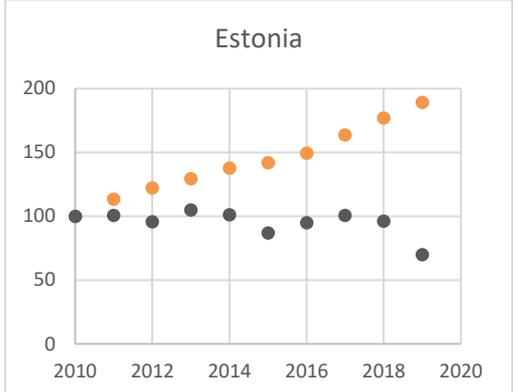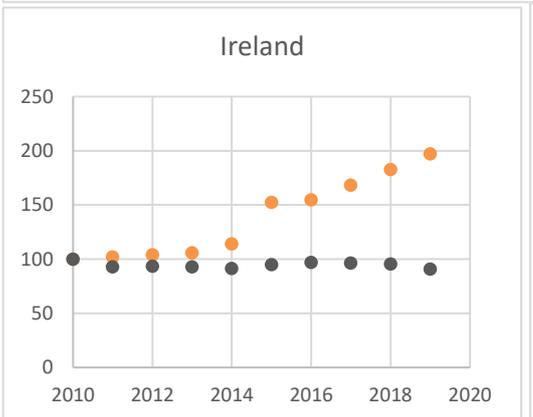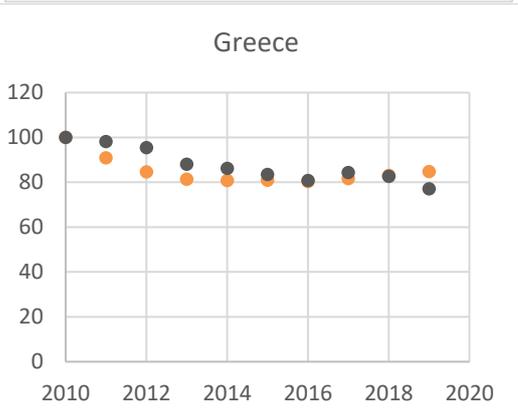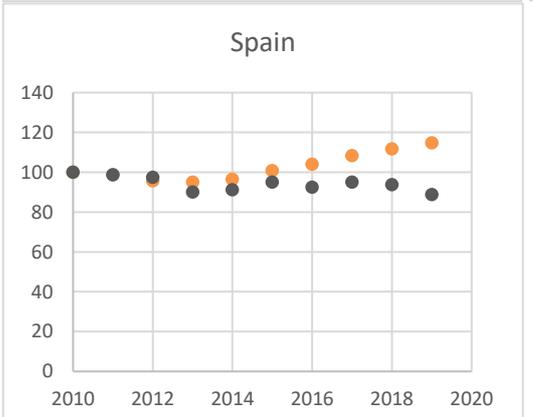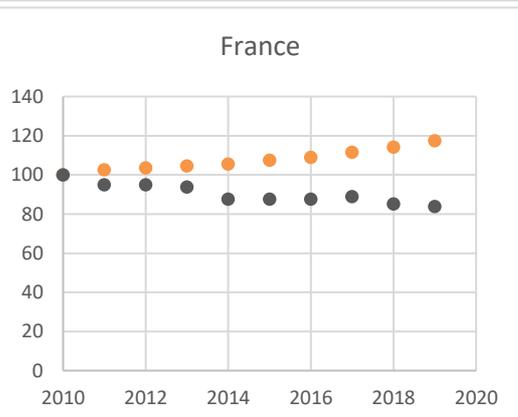

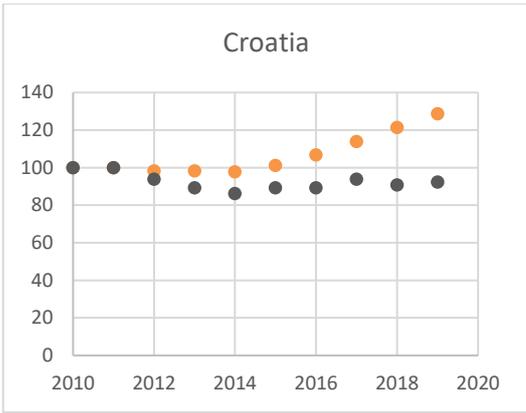
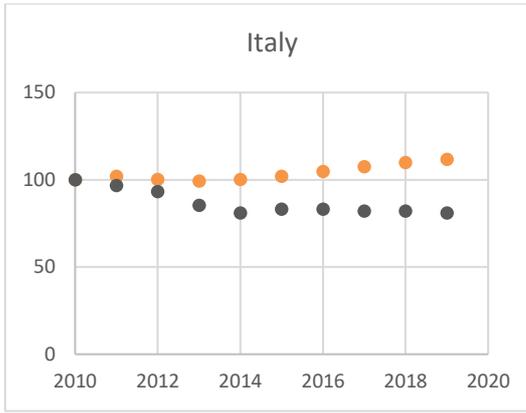
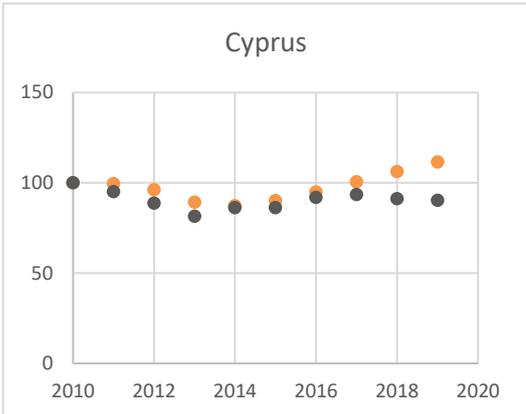
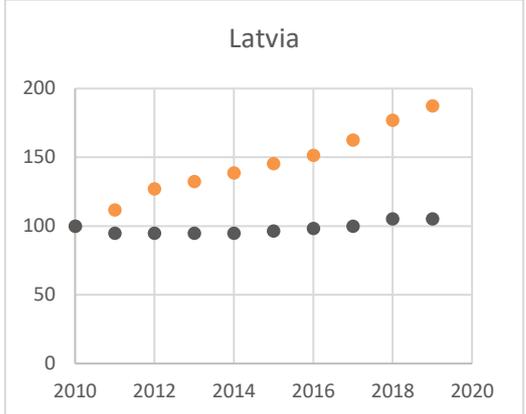
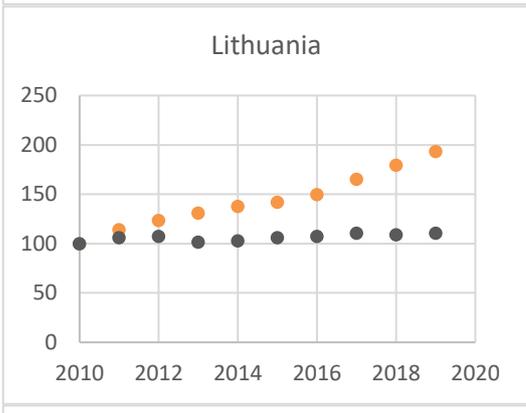
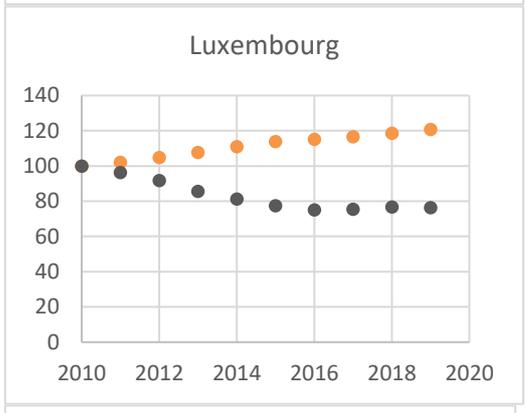
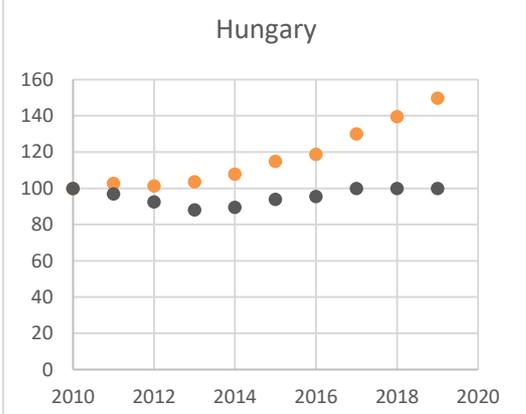
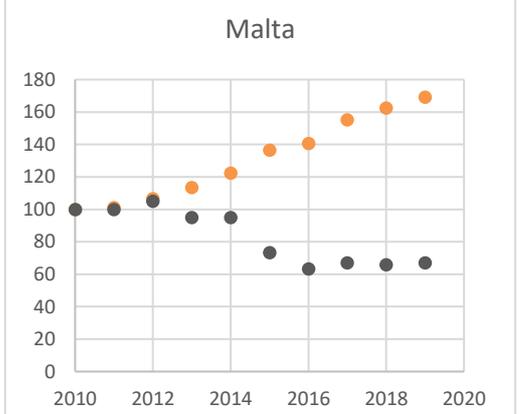

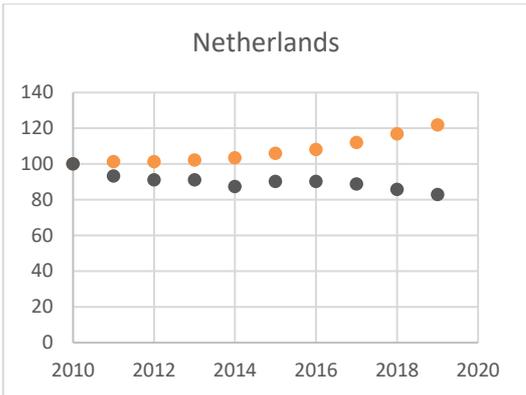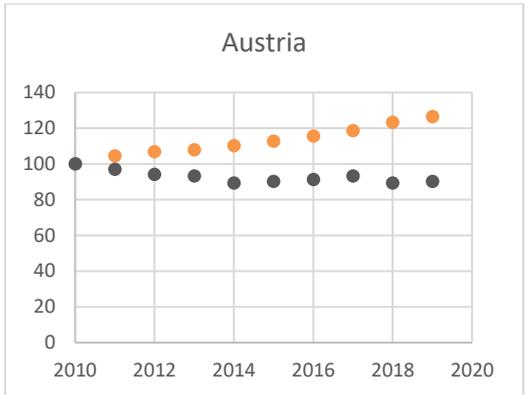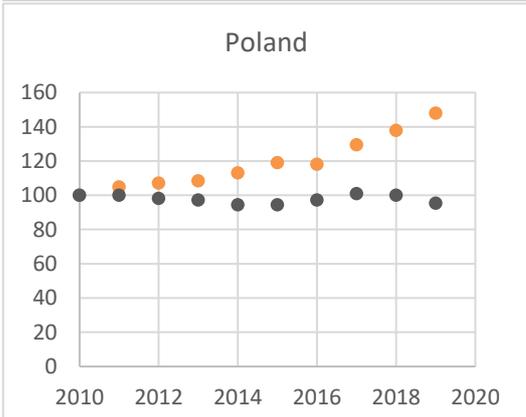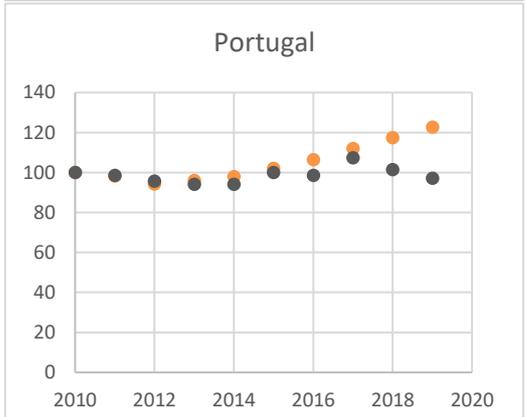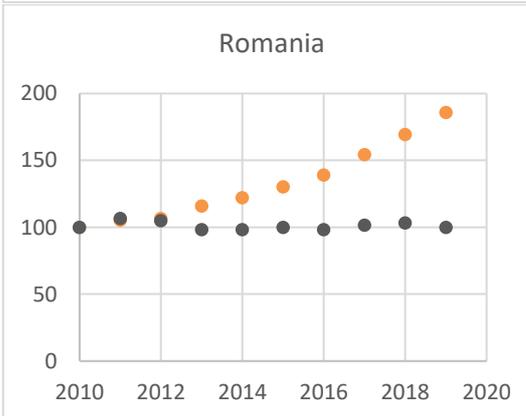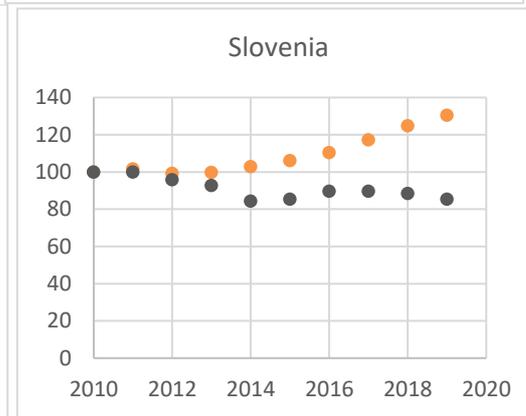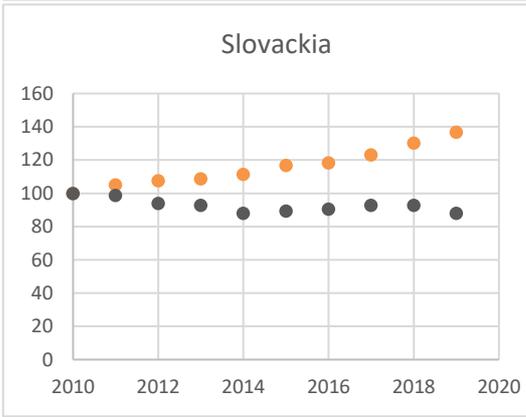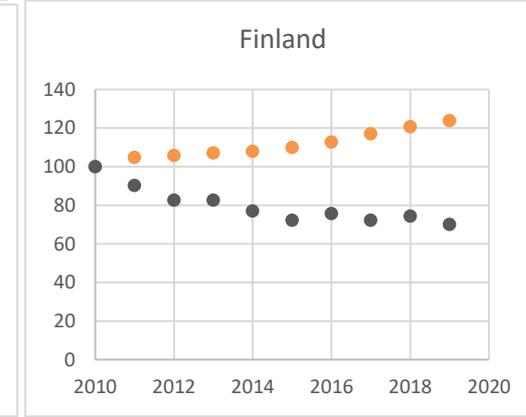

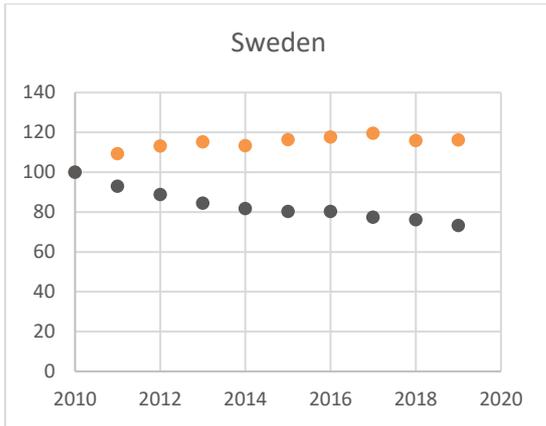

**Figure 4.** EU27 Member States economic Decoupling from the year 2010, set as reference = 100 for both GDP capita and GHG capita.

Figure 4 shows most of the EU 27 countries exhibit good decoupling according to the definition, even though the degree of decoupling can vary largely. Thus, in assessing the capability to pay for mitigation just looking at GDP per capita can be misleading. It is also very important to consider how GDP increases for each Member State correlate with GHG emissions: with less decoupling, a larger carbon budget may be necessary to invest in decarbonization [40] [41]. To transition energy and other economic sectors away from fossil fuels, it may be necessary to rely on those fossil fuels, at least temporally, provided that the burnt budget is spent to implement and scale up renewable resources and energy efficiency measures, objectives that are not still yet sufficiently pursued by Member States [42]. Thus, the Member States effort highlighted in Figure 4 should be monitored in future, to verify whether it is the result of the initial "efficiency measures" of business as usual economy rather than a decoupling due to renewables [43] and the circular economy implementation [44].

The carbon quotas due only to "Decoupling" are evaluated with Eq. 4, reported in Table A9 and shown in Figure 5.

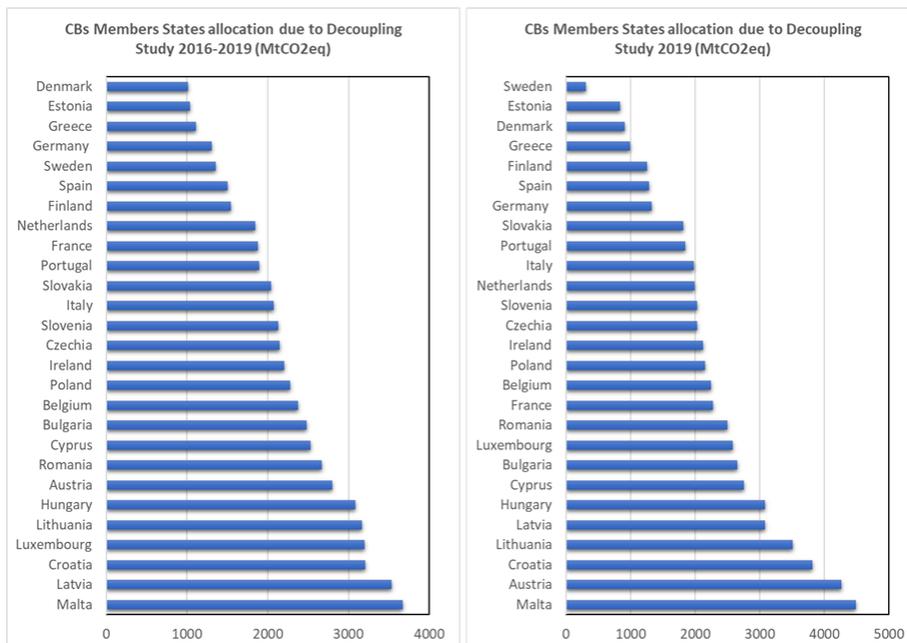

**Figure 5**. Member States CBs allocation according to the Decoupling Index by the authors. evaluated within Study 2016-2019 (left) and Study 2019 (right).

We can note that, while for CBs based on Capability (Figure 4) Member States allocation maintains the same ranking order in both the two studies, here, with a split based on decoupling, Carbon budgets magnitude and Member States ranking may vary significantly, at least for the several Member States.

*Blended approach Capability-Decoupling*

The previous findings led the authors to investigate a new carbon budget quota sharing we called "Economic term", which accounts for a blended "Capability" and "Decoupling" approach, according to Eq. 7.

This approach allows countries that have a less decoupled economy to have a higher amount of CB compared to the one only based on GDP per capita and vice versa. For those countries with a large decoupling, their capability carbon budget is reduced. Table 3 reports the carbon budget blending according to Eq.7 for the Study 2016-2019. Similar results are obtained in Study 2019 (see Table A.11). The use of w parameter, which spans from 0 to 1 with a step of 0.1, allows for blending the two contributions: with w=0 meaning only Capability criterion is considered and w= 1 only Decoupling criterion is considered; w=0.5 results in a carbon budget quota influenced in equal weight by the 2 indexes. Results are shown in Figure 6.

**Table 3**. Member States Carbon Budgets assignment with a different degree of a blend between Capability and Decoupling for Study 2016-2019.

|  | Capability | Capability- Decoupling blended Carbon Budget quotas (MtCo2eq) | | | | | | | | | Decoupling |
|---|---|---|---|---|---|---|---|---|---|---|---|
| **Belgium** | 1170 | 1291 | 1411 | 1531 | 1652 | 1772 | 1892 | 2013 | 2133 | 2253 | 2374 |
| **Bulgaria** | 5986 | 5636 | 5285 | 4935 | 4585 | 4234 | 3884 | 3534 | 3183 | 2833 | 2482 |
| **Czechia** | 2445 | 2414 | 2384 | 2354 | 2324 | 2294 | 2264 | 2234 | 2204 | 2173 | 2143 |
| **Denmark** | 903 | 913 | 924 | 935 | 945 | 956 | 967 | 977 | 988 | 998 | 1009 |
| **Germany** | 1163 | 1177 | 1192 | 1207 | 1221 | 1236 | 1250 | 1265 | 1280 | 1294 | 1309 |
| **Estonia** | 2476 | 2332 | 2188 | 2044 | 1900 | 1756 | 1612 | 1468 | 1324 | 1180 | 1036 |
| **Ireland** | 721 | 869 | 1017 | 1166 | 1314 | 1462 | 1610 | 1759 | 1907 | 2055 | 2204 |
| **Greece*** | 2799 | 2631 | 2462 | 2293 | 2125 | 1956 | 1787 | 1618 | 1450 | 1281 | 1112 |
| **Spain** | 1840 | 1806 | 1772 | 1739 | 1705 | 1671 | 1637 | 1603 | 1569 | 1536 | 1502 |
| **France** | 1341 | 1395 | 1449 | 1503 | 1557 | 1611 | 1665 | 1719 | 1773 | 1827 | 1881 |
| **Croatia** | 3725 | 3674 | 3622 | 3571 | 3519 | 3467 | 3416 | 3364 | 3312 | 3261 | 3209 |
| **Italy** | 1593 | 1641 | 1688 | 1736 | 1784 | 1832 | 1880 | 1927 | 1975 | 2023 | 2071 |
| **Cyprus** | 1924 | 1985 | 2045 | 2106 | 2167 | 2228 | 2289 | 2350 | 2410 | 2471 | 2532 |
| **Latvia** | 3208 | 3240 | 3273 | 3305 | 3337 | 3370 | 3402 | 3435 | 3467 | 3500 | 3532 |
| **Lithuania** | 2989 | 3007 | 3024 | 3042 | 3060 | 3078 | 3096 | 3113 | 3131 | 3149 | 3167 |
| **Luxembourg** | 473 | 745 | 1018 | 1291 | 1564 | 1836 | 2109 | 2382 | 2654 | 2927 | 3200 |
| **Hungary** | 3465 | 3426 | 3388 | 3350 | 3311 | 3273 | 3235 | 3196 | 3158 | 3120 | 3082 |
| **Malta** | 1803 | 1990 | 2177 | 2364 | 2551 | 2738 | 2925 | 3112 | 3299 | 3485 | 3672 |
| **Netherlands** | 1054 | 1133 | 1213 | 1292 | 1371 | 1451 | 1530 | 1609 | 1688 | 1768 | 1847 |
| **Austria** | 1086 | 1257 | 1429 | 1600 | 1772 | 1943 | 2115 | 2286 | 2458 | 2629 | 2800 |
| **Poland** | 3711 | 3567 | 3423 | 3280 | 3136 | 2992 | 2848 | 2705 | 2561 | 2417 | 2274 |
| **Portugal** | 2389 | 2339 | 2289 | 2240 | 2190 | 2140 | 2090 | 2040 | 1990 | 1940 | 1890 |
| **Romania** | 4625 | 4429 | 4234 | 4039 | 3844 | 3649 | 3453 | 3258 | 3063 | 2868 | 2672 |
| **Slovenia** | 2170 | 2166 | 2162 | 2157 | 2153 | 2149 | 2144 | 2140 | 2136 | 2131 | 2127 |
| **Slovakia** | 2902 | 2815 | 2729 | 2643 | 2557 | 2471 | 2385 | 2299 | 2212 | 2126 | 2040 |
| **Finland** | 1118 | 1161 | 1203 | 1246 | 1289 | 1332 | 1375 | 1417 | 1460 | 1503 | 1546 |
| **Sweden** | 993 | 1029 | 1066 | 1102 | 1138 | 1175 | 1211 | 1248 | 1284 | 1320 | 1357 |
| **w** | **0** | **0,1** | **0,2** | **0,3** | **0,4** | **0,5** | **0,6** | **0,7** | **0,8** | **0,9** | **1** |

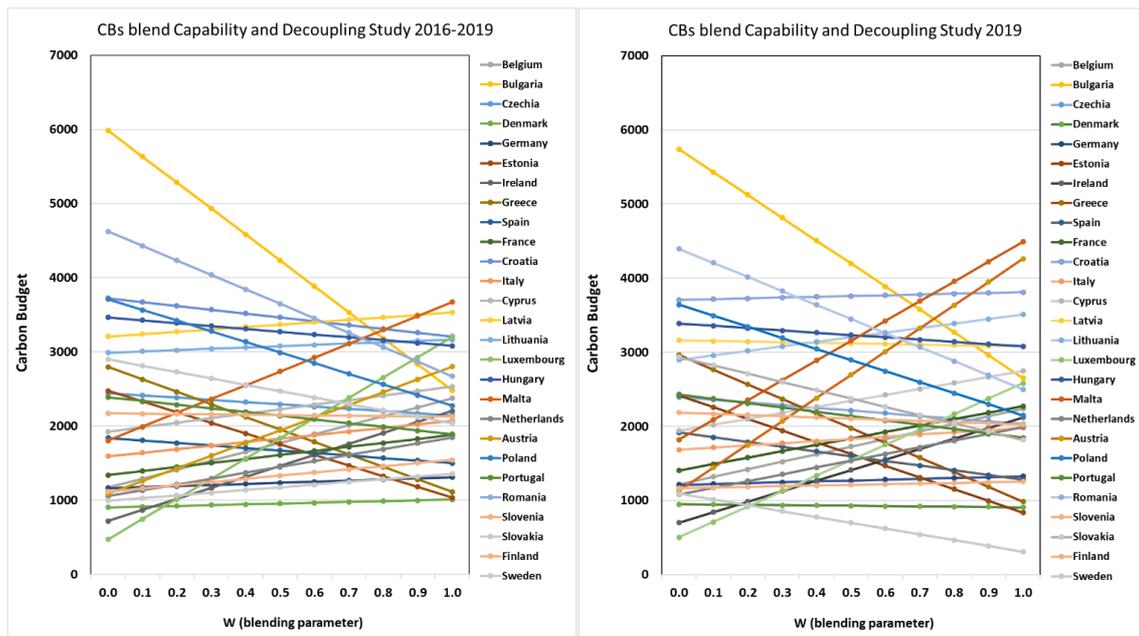

**Figure 6.** Member States CBs distribution according to Capability and Decoupling blended approach. Left: Study 2016-2019 data Table 3; right Study 2019, data in Table A.11.

Figure 6 shows how, for several countries, the carbon budget allocation can change based on the "carbonization" of GDP, measured using the decoupling terms. The effect of decoupling is to narrow the carbon budget range between 1000 - 4000 MtCOeq in Study 2016-2019 and between 500-4500 in Study 2019. This is a reasonable result as we have shown before (Figure 4) given that the Member States rescaled index ranges from -1 to -4 so CBs redistribution only based on Decoupling results in a narrower range. This result also highlights the importance of reconsidering the concept of economic "capability to invest for mitigation" only based on a GDP measurement, with a Capability-Decoupling blended approach being more appropriate.

However, to complete reliable mathematical modelling of carbon budget distribution across the EU Member States, the Inertia term must be considered. This is particularly evident looking at Germany, Spain, France and Italy's carbon budget estimates according to Capability or Decoupling (or even in the blend of them) in comparison to their carbon budget allocated for the ESR sector only until 2030 (Table A.4): each of these countries have smaller carbon budgets according to the capability-decoupling criteria to 2050 than the ones allocated to ESR by 2030. Splitting the EU27 budget based only on the Economic term would dramatically lower carbon budgets for those countries creating higher costs which may not be feasible. On the other side, countries such as Malta, Slovenia, Cyprus and others, which historically have lower emissions, end up with a carbon budget which is too large. This means that an approach based solely on capability and decoupling, or solely on Inertia, are both insufficient to assess a reliable distribution of the budgets, thus we approached the blend of the three.

***Blended approach Inertia-Capability-Decoupling***

Member States' carbon budget allocation due to the Inertia term in Eq.6 is reported in Table A.10 and summarized in Figure 7, for the Study 2016-2019 and the Study 2019.

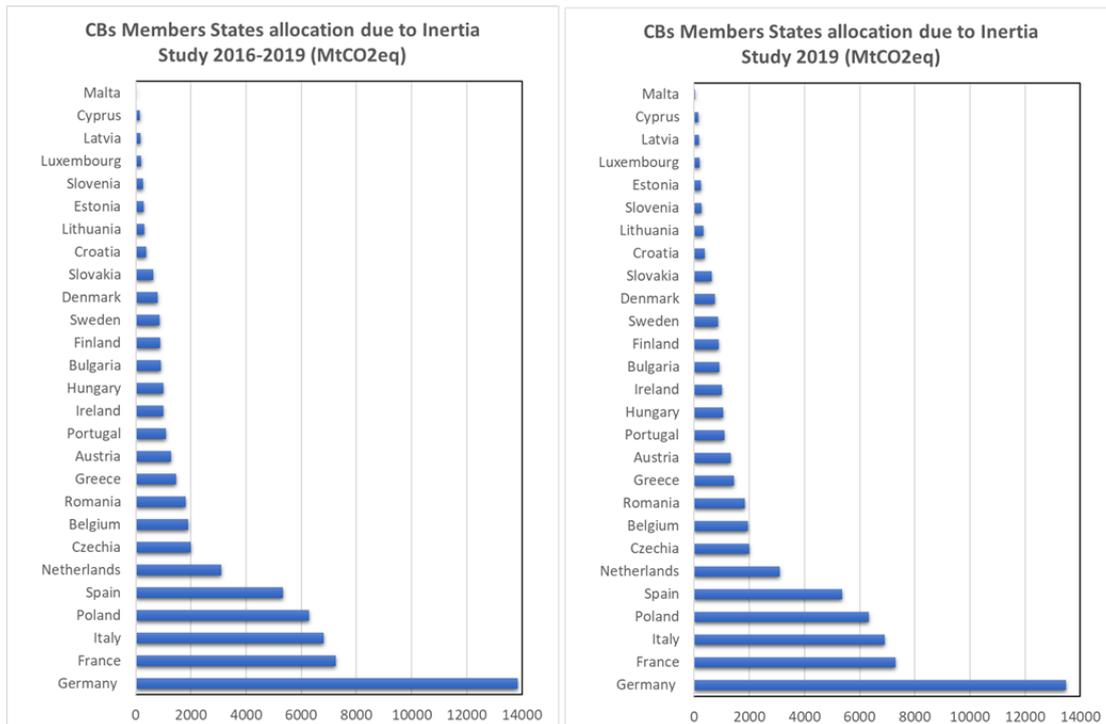

**Figure 1**. Member States CBs allocation according to the Inertia Principle

As previously commented, Inertia has a relevant weight in distributing quotas for those countries that have a larger capability, but which is due to higher emission rates (Germany, Spain, France and Italy for instance), or that, vice versa, show low emission rates, but little or no decoupling which then resulted in a higher carbon budget than necessary (Malta, Cyprus, Slovenia and other). We remember here that the Inertia term is added to the Eq. 7 to become Eq. 8.

For comparison, especially regarding the effect of decoupling in the economic part of the blend (named Capability plus Decoupling), discussed at 3.4, a first blend has been created including only Inertia (Eq.6) and Capability (Eq.4) with a parameter "z" spanning between 0 and 1 (step 0.1), exactly as we did for blending Capability and Decoupling. z=0 means only Capability is considered to split the EU27 Carbon Budget among Member States; z=1 means only Inertia is considered to split the EU27 Carbon Budget among member States. This is the approach usually considered in assigning a carbon budget quota based on Effort Sharing [30]. Subsequently, a second blend including all 3 terms (Inertia, Capability and Decoupling) has been estimated: as for Eq. 8, Inertia is blended with the Economic term which includes Capability and Decoupling that are, in turn, blended between themselves by w.

*Results for Study 2016-2019*

Results of the EU27 Carbon Budget split by means of the Capability Inertia blend and the Inertia Capability and Decoupling blend are in Figure 8.

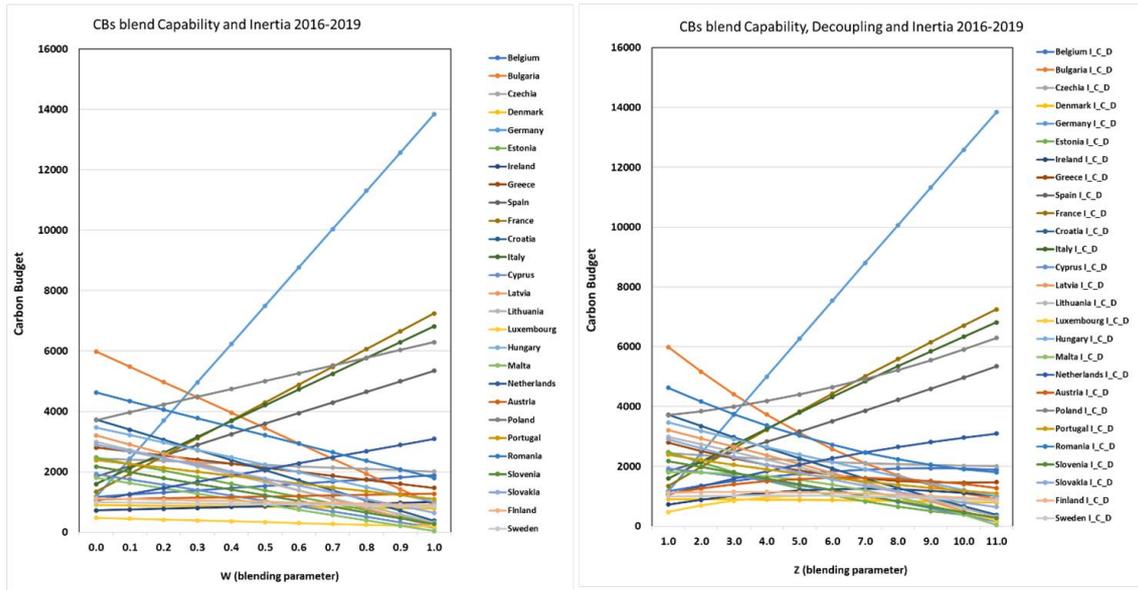

**Figure 8.** Assessing Member States quotas of EU27 Carbon Budget. Left, Capability – Inertia blended approach, data Table A.12; right, Inertia – Capability – Decoupling (I_C_D) data Table A.14 for Study 2016-2019.

When using the three terms, some of the Member States' carbon budget blended trends may experience changes in the shapes, from straight (Figure 8 left) to concave and/or convex curves (Figure 8 right); others do not change and remain almost straight.

For w between 0.1 to 0.9, four groups of behaviours are detectable. In Group 1 (Figure 9) the CB trends show very minor changes when introducing or not the decoupling in the Capability-Inertia blend (straight lines). In Group 2 (Figure 10), all blended CBs are between Capability (z=0) and Inertia (z=1) term CBs, but lower (concave curves) than the ones without decoupling. In Group 3 (Figure 11) blended CBs are between Capability and Inertia term CBs but are higher (convex curves) than ones without decoupling. In Group 4 (Figure 12) blended CBs may take values higher (pronounced convex) than capability (z=0) and inertia terms (z=1). Those changes in the shape of the curves can be explained in relation with the Tapio index and the gap between the Inertia budget (Figure 7) and the Capability budget (Figure 3). This gap was estimated as a percentage difference of the Inertia budget versus Capability budget (how much larger or smaller is the Inertia budget in comparison to the Capability); data are reported in Table 4. In the same table the Tapio indexes are reported.

**Table 4.** Member States Capability vs Inertia CBs gap and Tapio index for Study 2016-2019.

| Countries | CBs Capability vs Inertia gap (%) | Tapio Index |
|---|---|---|
| **Germany** | < -100% | -1.554 |
| **France** | <-100% | -0.701 |
| **Italy** | < -100% | -0.522 |
| **Netherlands** | <- 100% | -0.737 |
| **Spain** | <-100% | -1.194 |
| **Poland** | -69.58% | -0.364 |
| **Belgium** | -61.54% | -0.296 |
| **Ireland** | -38.96% | -0.415 |
| **Austria** | -17.37% | -0.060 |
| **Denmark** | 12.98% | -2.388 |
| **Sweden** | 14.19% | -1.456 |

| Czechia | 18.02% | -0.462 |
|---|---|---|
| Finland | 19.87% | -1.125 |
| Greece | 47.57% | -2.050 |
| Portugal | 54.14% | -0.691 |
| Luxembourg | 60.27% | 0.104 |
| Romania | 61.12% | -0.123 |
| Hungary | 71.10% | 0.060 |
| Slovakia | 77.85% | -0.549 |
| Bulgaria | 84.76% | -0.228 |
| Slovenia | 87.50% | -0.475 |
| Estonia | 88.06% | -2.292 |
| Lithuania | 89.31% | 0.092 |
| Croatia | 89.85% | 0.107 |
| Cyprus | 92.08% | -0.199 |
| Latvia | 94.50% | 0.212 |
| Malta | 97.86% | 0.251 |

In Group 1, we can see the Inertia term dominates whether it is much higher or much lower than the capability term. We note that CBs assessed with the Capability term are proportionally inverse to the GDP, thus, as previously discussed for countries like Germany, France, Spain and the others within this group, because of the high GDP, CBs cannot be reduced a lot, even when those countries show a strong decoupling index (see Table 4). Similarly for those countries like Latvia, Lithuania and Croatia, which have a smaller capability to pay and a weak decoupled index, the CBs from capability might result in being too high, as they are very low-emitter countries.

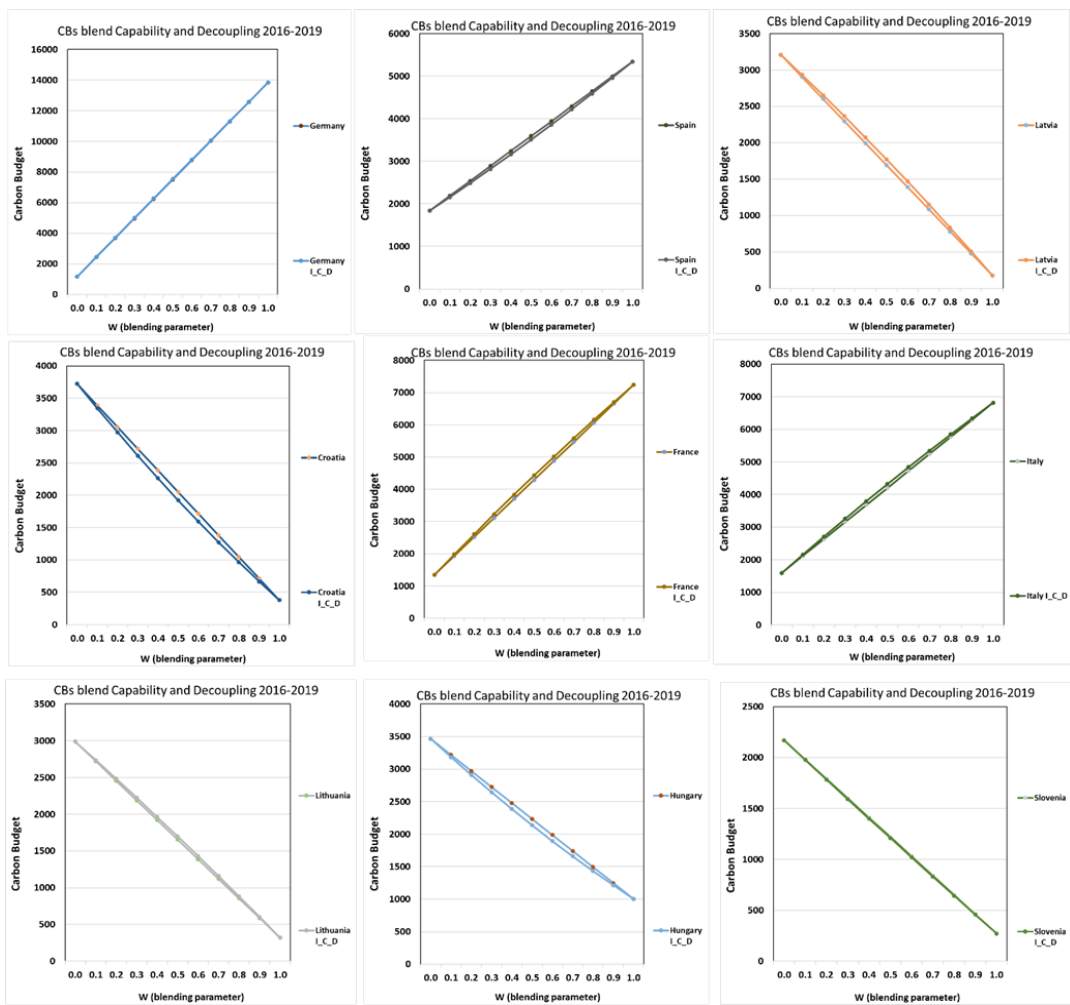

**Figure 9.** Member States in Group 1, Inertia term dominates.

In Group 2 (Figure 10) the difference between Capability and Inertia CBs is smaller than in Group 1, and almost all those countries show a quite strong decoupling index whose effect is to suggest that blended CBs for those countries can be smaller than the ones evaluated only considering their Capability and Inertia.

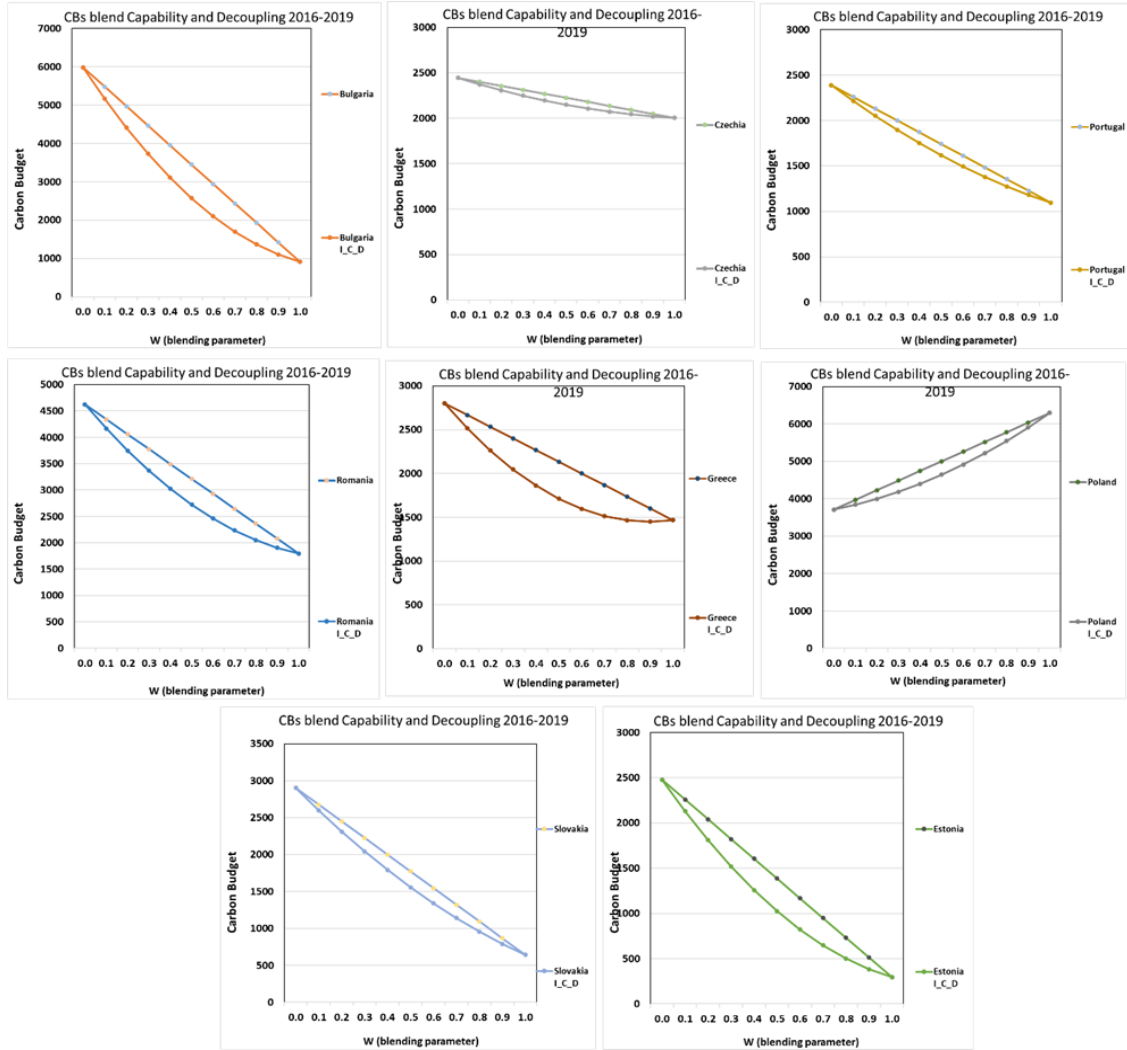

**Figure 10.** Member States in the Group 2, high decoupling term dominates.

In Group 3, we have 2 countries, Cyprus and Malta where CBs due to capability are much higher than emission rate, as that in the first group, but here the decoupling is weak, this is the reason for an increasing carbon budget in the blend. As for Group 1, reliable CBs should be closer to the CBs due to the Inertia term, with the low degree of decoupling suggesting they need to be a bit higher due to the carbonization of those economies (weak decoupling). Denmark shows similar behaviour but with opposite findings: the difference between Capability and Inertia CB is low and this makes the contribution of carbon budget based on decoupling quota emerge more strongly. This case will be discussed further within Group 4. The Netherlands behaviour is instead more similar to the ones in Group 1, even the Capability Inertia gap has a minor influence, and the effect of decoupling is more visible than for Germany, Spain, France and Italy.

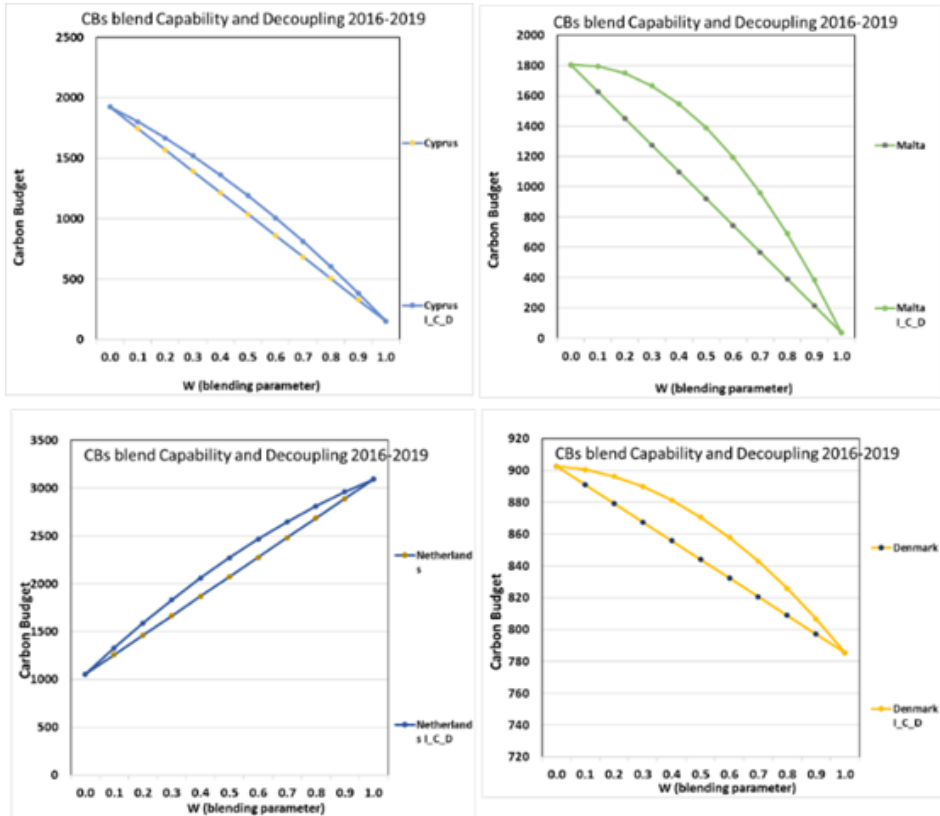

**Figure 11.** Member States in the Group 3, weak decoupling term dominates.

Group 4 is the most sensitive to the introduction of decoupling: each of these Member States shows CBs assessed with the decoupling index larger than that from using Inertia and Capability. As expected, the contribution of Decoupling increased with the decrease of the Tapio index.

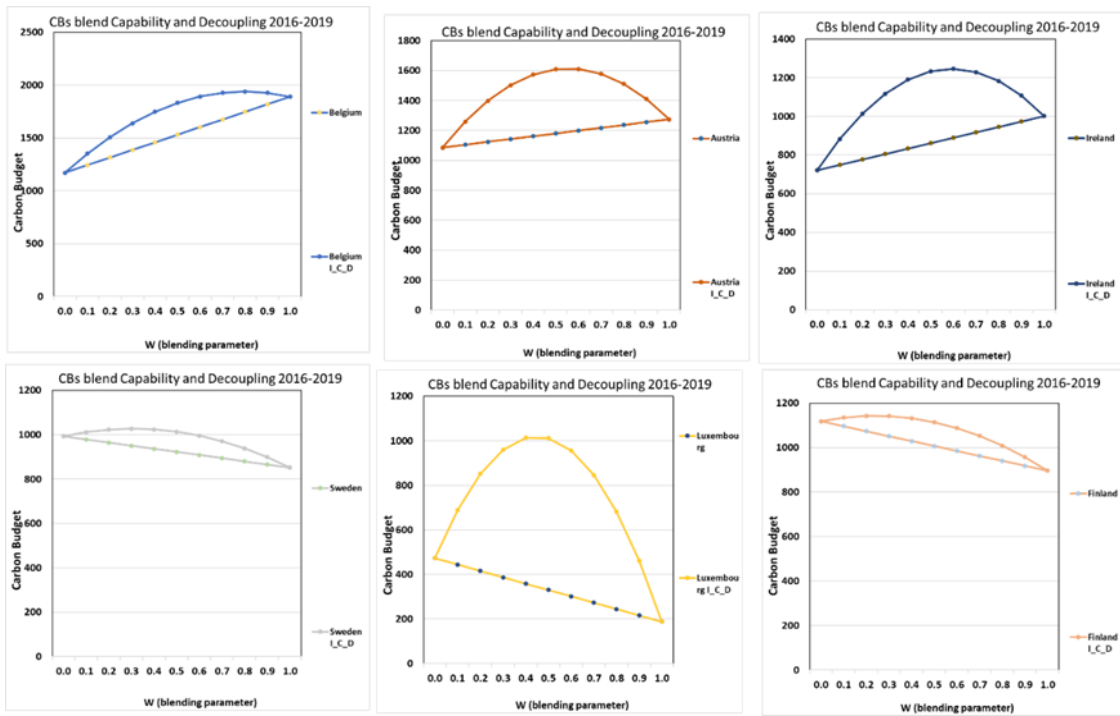

**Figure 12.** Member States in Group 4: low Capability-Inertia gap, makes emerge decoupling effects.

*Results for Study 2019*

In Study 2019, the same approach of Study 2016-2019 has been repeated considering that, for EU27 as a whole, the Tapio index is equal to -1.02 and higher than the one in the 2016-2019 interval (equal to 0.75, see Table 2). This analysis aims to assess the most recent position of the Members States on the pathway to decarbonization in comparison to the one described as average during the 2016-2019 years. The results are in Figure 13.

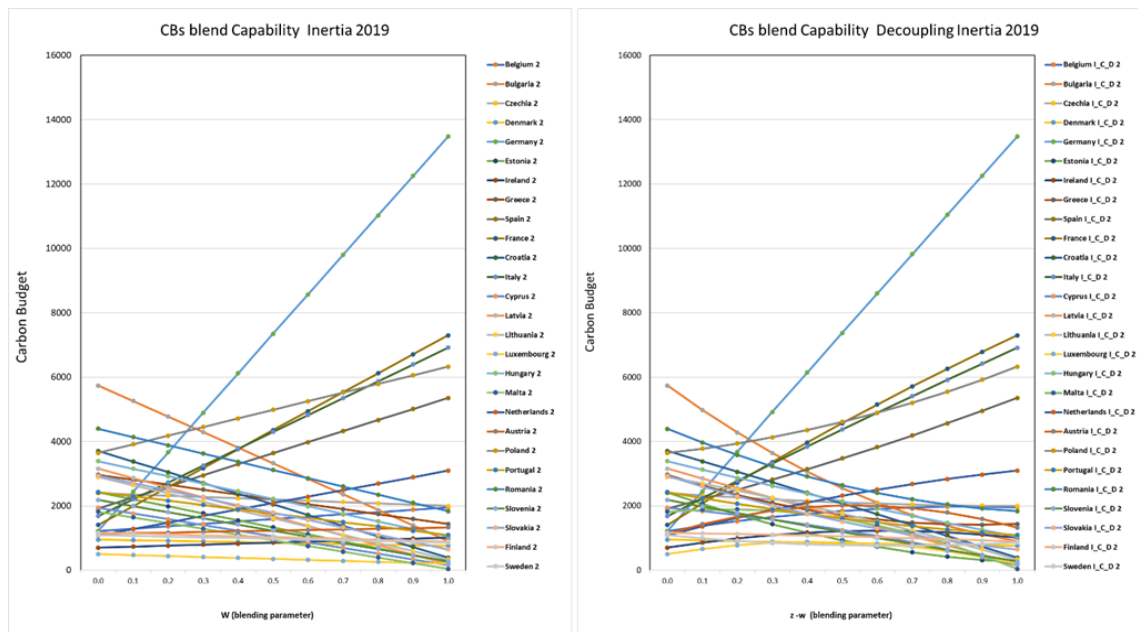

**Figure 13.** Assessing Member States quotas of EU27 Carbon Budget left, Inertia-Capability blended approach, data Table A.13; right, Inertia – Capability – Decoupling (I_C_D 2) data Table A.15 for Study 2019.

Member States in Group 1, with the Inertia dominating, are the same in Study 2016-2019 and the Study 2019.

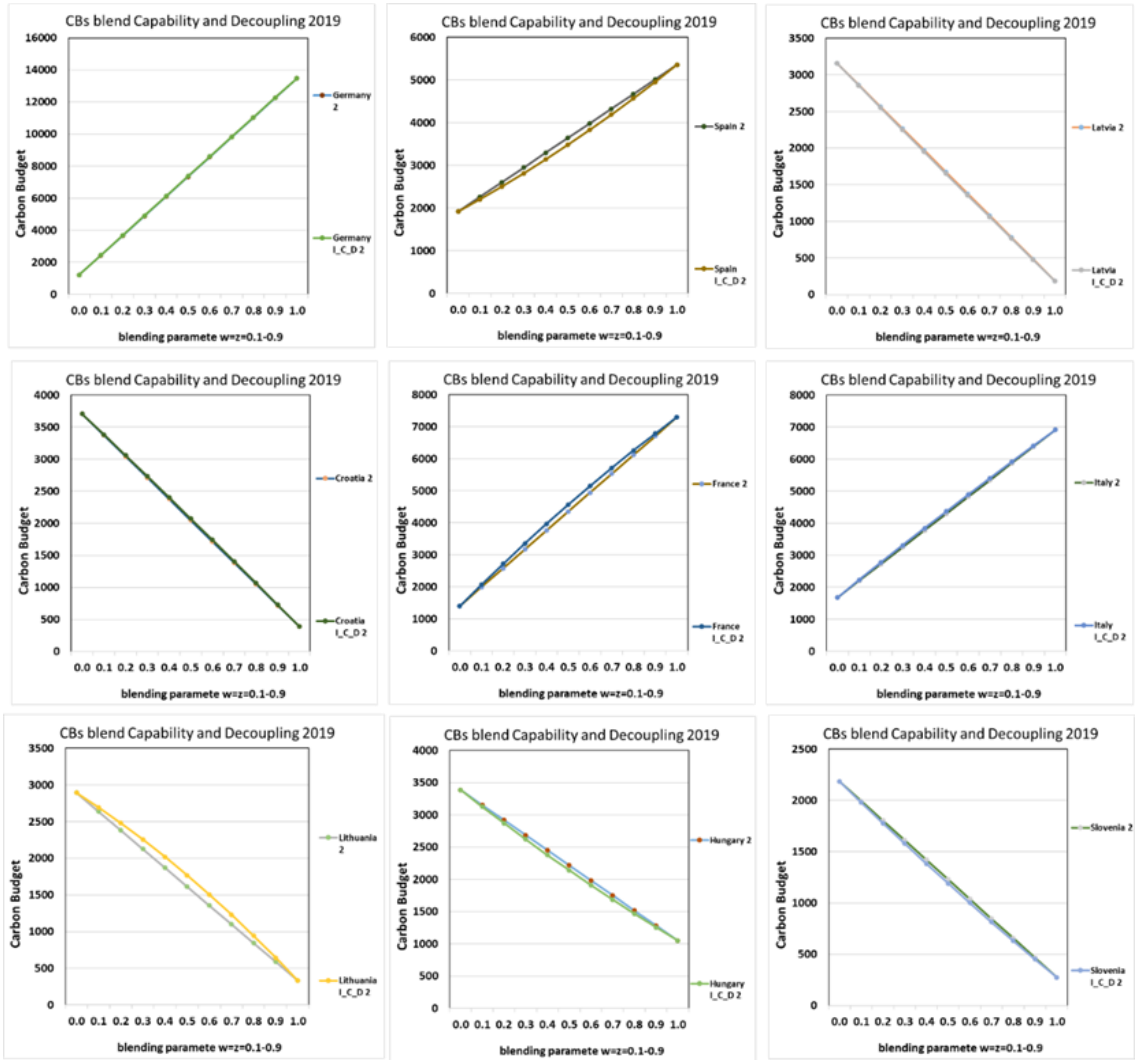

**Figure 14.** Member States in the Group 1(2019): Inertia term dominates, and some minor decoupling effects are visible for Member states decoupling index get worse or better than in 2016-2019.

Note however that for countries like Lithuania and France, a small contribution from the decoupling appears (convex curves), because both countries have a slightly worse Tapio Index in comparison to 2016-2019; vice versa for Spain, whose Tapio index gets better (small concave) (Table 5).

**Table 5.** Member States Capability vs Inertia CBs gap and Tapio index for Study 2019.

| Countries | CBs Capability vs Inertia gap (%) | Tapio Index |
|---|---|---|
| **Germany** | < -100% | -1.93 |
| **France** | < -100% | -0.52 |
| **Italy** | < -100% | -0.81 |
| **Netherlands** | < -100% | -0.80 |
| **Spain** | < -100% | -2.05 |
| **Poland** | -74% | -0.63 |
| **Belgium** | -61% | -0.55 |
| **Ireland** | -45% | -0.66 |
| **Austria** | -17% | 0.41 |
| **Czechia** | 17% | -0.75 |
| **Denmark** | 20% | -3.49 |

| Sweden | 21% | -13.18 |
| Finland | 23% | -2.12 |
| Greece | 51% | -3.10 |
| Portugal | 55% | -0.97 |
| Romania | 58% | -0.34 |
| Luxembourg | 60% | -0.28 |
| Hungary | 69% | 0.00 |
| Slovakia | 78% | -1.01 |
| Bulgaria | 84% | -0.24 |
| Slovenia | 87% | -0.76 |
| Lithuania | 89% | 0.18 |
| Croatia | 90% | 0.28 |
| Estonia | 90% | -3.92 |
| Cyprus | 92% | -0.18 |
| Latvia | 94% | 0.00 |
| Malta | 98% | 0.46 |

Also, the Member States in Group 2 (2019) (Figure 15) are the same in the Study 2016-2019 (Figure 14) confirming that for all these countries' CBs can be assessed as a bit lower than the CBs blend due to Capability and Inertia.

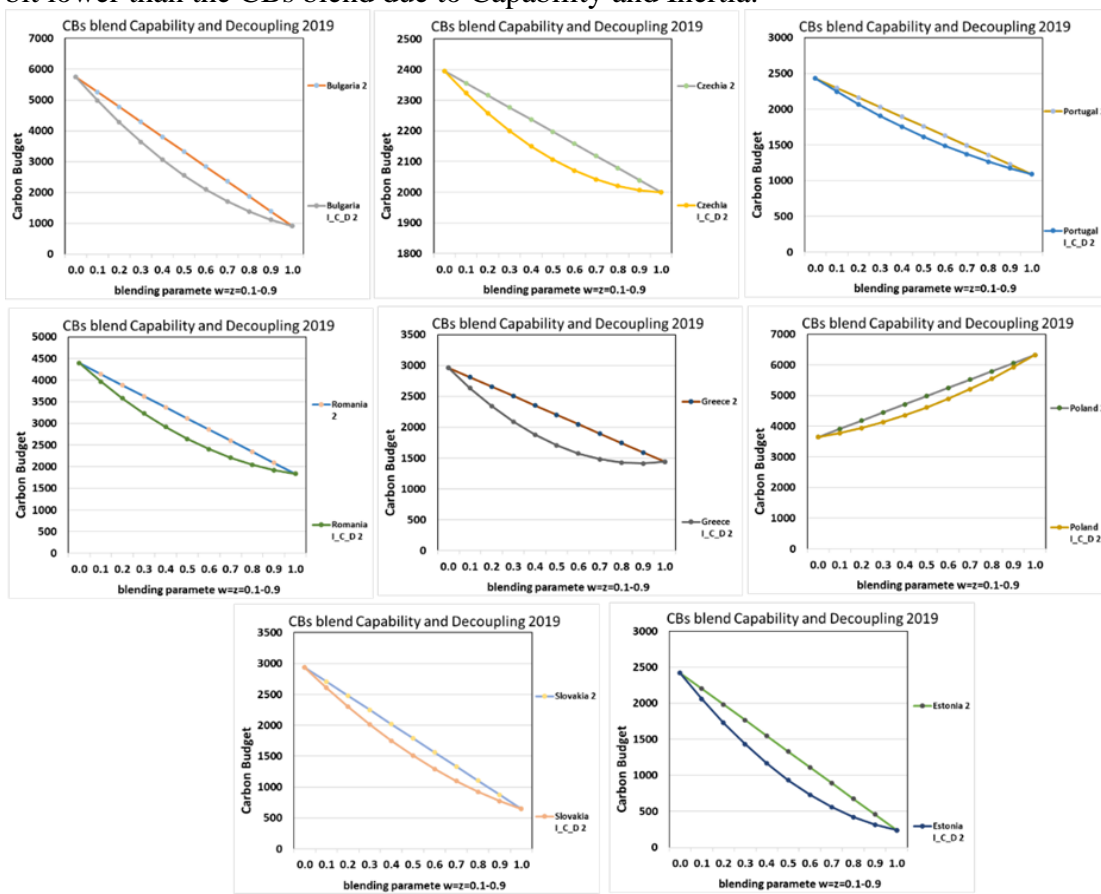

**Figure 15.** Member States in the Group 2 (2019), high decoupling term dominates, are the same as the Study 2016-2019 even with different degrees of decoupling.

Cyprus, Netherlands, and Denmark are still in Group 3 (Figure 16) and show the same trend as that in the Study 2016-2019, with different intensities of decoupling. We found now that Finland joins this group while Malta moves to Group 4. Finland further decoupled its economy, inertia is less, so that the gap capability-inertia has reduced, and a very small decoupling effect is seen.

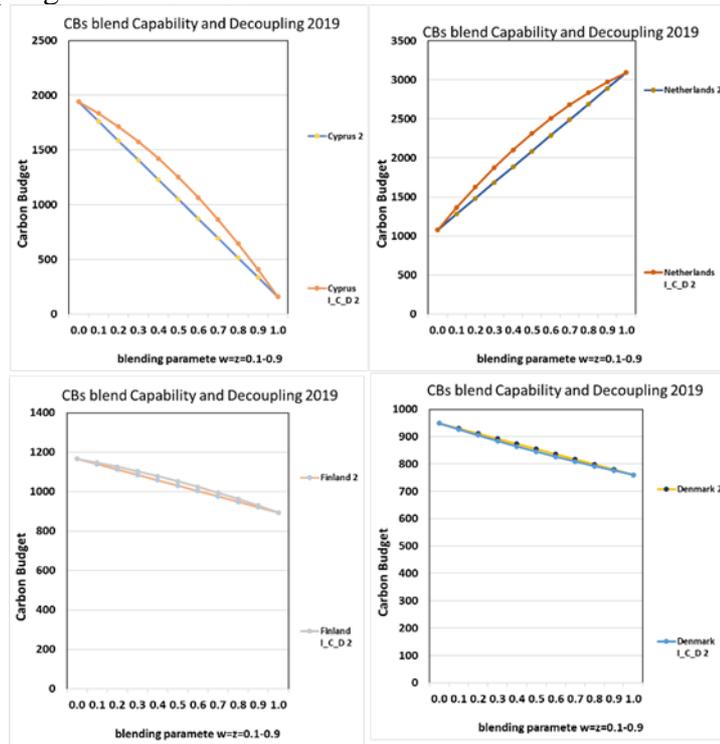

**Figure 16.** Member States in the Group 3 (2019), high decoupling term dominates, are the same of the Study

Finally Group 4 (Figure 17) refers to countries that are more sensitive to the decoupling effect. It is particularly important to note the high effect of decoupling on Sweden, whose decoupling in 2019 is the highest of the EU27, so that its blended carbon budget results are much lower than considering Capability and Inertia only.

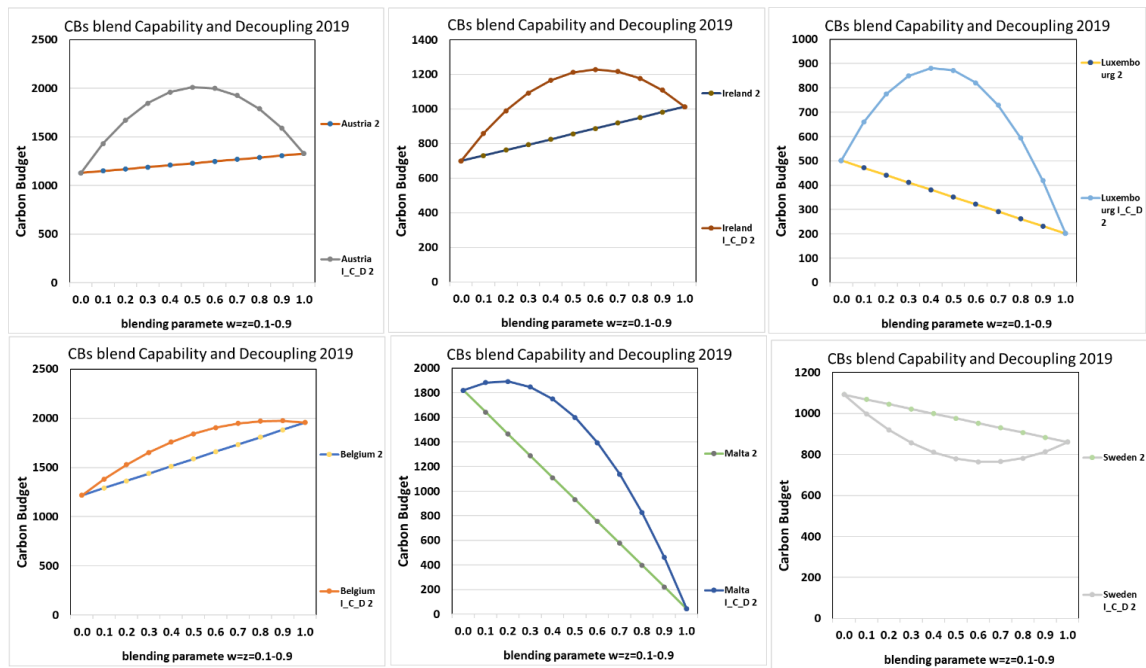

**Figure 17**. Member States in the Group 4 (2019), low Capability-Inertia gap, makes emerge decoupling effects

*Carbon budget allocation for the EU27 member States*

In Study 2016-2019 and Study 2019 we have created and analysed a set of CB trends according to Capability, Decoupling, and Inertia principles. From the analysis we find that the distribution of those budgets is still mainly driven by the Inertia principle. Despite their capability to pay and even their degree of decoupling, the EU big emitters from one side and lower emitters from the other, frame the whole distribution. It is not possible, at the moment, to introduce an effective capability criterion as a main driver in assessing carbon quotas for Member States, even a degree of decoupling can work as a modulator and raise or lower the Inertia term on the basis of Inertia Capability gap and the Tapio index. This is evident when comparing CBs from the ESR until 2030 with the ones reported for the Capability, Decoupling, and Inertia blend of Study 2016-2019 (Table 6). Looking at Figure 2, the ETS CBs to 2030 will account for slightly less budget than the ESR to 2030. The total budget is equivalent to the 2030 ESR quota multiplied by 3 approximately, and a similar proportion can be adopted to focus (bold cells, table 6) a range of suitable CBs for each country (country budget as 3 times their present budget for ESR to 2030, which are reported in the grey column of table 6). This estimation is not accurate for some countries (Bulgaria, Czechia, Denmark and others) that show all blended budgets higher than 3 times the ESR one; these results indicate that the distribution is inertia driven by the higher emitters and a range of z between 0.5-1 might represent the more suitable blends. Similar considerations can be made for Study 2019.

**Table 6.** Comparison of 2021-2030 CBs from ESR (grey shadow) with 2021-2050 CBs from the blended approach evaluated in the Study 2016-2019. Bold cells indicate budgets that are at least 3 times higher than the ESR budget reported in the grey-shadowed column, which is the ESR budget to 2030.

| | Capability | Capability- Decoupling-Inertia blended CBs Study 2016-2019 (MtCo2eq) | | | | | | | | | Inertia | CBs from ESR 2021-2030 |
|---|---|---|---|---|---|---|---|---|---|---|---|---|
| **Belgium** | 1170 | 1242 | 1314 | 1386 | 1458 | 1530 | 1603 | **1675** | **1747** | **1819** | **1891** | 554 |
| **Bulgaria** | 5986 | 5479 | 4971 | 4464 | 3957 | 3449 | 2942 | 2434 | 1927 | 1419 | 912 | 211 |
| **Czechia** | 2445 | 2400 | 2356 | 2312 | 2268 | 2224 | 2180 | 2136 | 2092 | 2048 | 2004 | 538 |
| **Denmark** | 903 | 891 | 879 | 868 | 856 | 844 | 832 | 821 | 809 | 797 | 785 | 250 |
| **Germany** | 1163 | 2430 | 3698 | 4966 | 6233 | 7501 | 8769 | **10036** | **11304** | **12572** | **13840** | 3258 |
| **Estonia** | 2476 | 2258 | 2040 | 1822 | 1604 | 1386 | 1168 | 950 | 732 | 514 | 296 | 54 |
| **Ireland** | 721 | 749 | 777 | 805 | 833 | 861 | 889 | 917 | 945 | 973 | **1002** | 347 |
| **Greece** | 2799 | 2666 | 2533 | 2400 | 2267 | 2134 | 2000 | 1867 | 1734 | 1601 | 1468 | 439 |
| **Spain** | 1840 | 2190 | 2541 | 2892 | 3242 | 3593 | 3943 | 4294 | 4644 | **4995** | **5346** | 1662 |
| **France** | 1341 | 1931 | 2522 | 3112 | 3703 | 4293 | 4883 | 5474 | 6064 | 6655 | 7245 | 2639 |
| **Croatia** | **3725** | 3391 | 3056 | 2721 | 2387 | 2052 | 1717 | 1382 | 1048 | 713 | 378 | 150 |
| **Italy** | 1593 | 2115 | 2637 | 3159 | 3681 | 4203 | 4725 | 5247 | 5769 | 6291 | **6814** | 2216 |
| **Cyprus** | 1924 | 1747 | 1570 | 1392 | 1215 | 1038 | 861 | 684 | 507 | 329 | 152 | 35 |
| **Latvia** | **3208** | 2905 | 2602 | 2298 | 1995 | 1692 | 1389 | 1086 | 783 | 480 | 177 | 82 |
| **Lithuania** | 2989 | 2722 | 2455 | 2188 | 1921 | 1654 | 1387 | 1120 | 853 | 587 | 320 | 124 |
| **Luxembourg** | 473 | 444 | 416 | 387 | 359 | 330 | 302 | 273 | 245 | 216 | 188 | 65 |
| **Hungary** | **3465** | 3218 | 2972 | 2726 | 2479 | 2233 | 1987 | 1740 | 1494 | **1248** | 1001 | 385 |
| **Malta** | 1803 | 1627 | 1450 | 1274 | 1097 | 921 | 744 | 568 | 392 | 215 | 39 | 11 |
| **Netherlands** | 1054 | 1258 | 1462 | 1666 | 1869 | 2073 | 2277 | **2481** | **2684** | **2888** | **3092** | 776 |
| **Austria** | 1086 | 1105 | 1124 | 1143 | 1161 | 1180 | 1199 | 1218 | 1237 | 1256 | 1275 | 378 |

| | | | | | | | | | | | |
|---|---|---|---|---|---|---|---|---|---|---|---|
| **Poland** | 3711 | 3969 | 4227 | 4485 | 4743 | 5002 | 5260 | 5518 | 5776 | 6034 | 6293 | 1855 |
| **Portugal** | 2389 | 2260 | 2131 | 2001 | 1872 | 1743 | 1613 | 1484 | 1354 | 1225 | 1096 | 353 |
| **Romania** | 4625 | 4342 | 4059 | 3777 | 3494 | 3211 | 2929 | 2646 | 2363 | 2081 | 1798 | 701 |
| **Slovenia** | 2170 | 1981 | 1791 | 1601 | 1411 | 1221 | 1031 | 841 | 651 | 461 | 271 | 87 |
| **Slovakia** | 2902 | 2676 | 2450 | 2224 | 1998 | 1772 | 1546 | 1320 | 1094 | 869 | 643 | 191 |
| **Finland** | 1118 | 1096 | 1074 | 1051 | 1029 | 1007 | 985 | 962 | 940 | 918 | 896 | 222 |
| **Sweden** | 993 | 979 | 965 | 951 | 937 | 923 | 908 | 894 | 880 | 866 | 852 | 250 |
| w | 0 | 0,1 | 0,2 | 0,3 | 0,4 | 0,5 | 0,6 | 0,7 | 0,8 | 0,9 | 1 | **Total 17834** |

Due to the simplicity of the blending model, this is just a first assessment; further adjustment is necessary to refine the assignment because of the further characteristics of each country. However, with this approach it is now possible to assess and review Member States' carbon budget apportions as frequently as necessary considering the progress or difficulties each Member State may face during the decarbonization pathways. The method helps to assess, for each Member State, the carbon budget that should not be exceeded or, in case of unexpected increases, to plan a compensation strategy with quotas from countries that are performing better. A similar portioning approach can also be used to further distribute quotas within ESR or ETS sectors or the main EU27 economics sectors. This will be part of further research.

**Conclusion**

Within the context of the Green Deal targets, and with the recent invasion of Ukraine pushing the EU to consider a more rapid move away from oil and gas (at least from Russian oil and gas), decarbonizing the European Union is more urgent than ever. Despite the European Union's pledges, the present study highlights that the Member States show a large variety of emission behaviours that are difficult to manage with common guidelines and approaches. To fill this gap, the authors propose a simple mathematical model which can allocate carbon quotas among the member states based on their historical emissions behaviour, their capability to spend on mitigation and the level of economic decoupling already achieved. As decarbonization is a dynamic process, this model allows for the quick redistribution of carbon quotas according to the progress or difficulties (expressed in terms of changes in emissions trends, GDP trends and decoupling trends) each Member State may face during the decarbonization pathways. Due to its simplicity, the model does not pretend to fully capture and manage a Member State's emissions features, it rather aims to be a recursive easy-to-use policy tool to monitor the EU27 decarbonization pathway.

**Acknowledgements**. This research was funded by PLEDGES project- European Union's Horizon 2020 MSCA IF GA ID: 101023109.
**Data availability statement.** The authors confirm that the data supporting the findings of this study are available within the article and its supplementary materials.

**Reference**

[1] Eurostat, "Population projections in the EU- Statistic explained," *Eurostat*, Sep. 2021. https://ec.europa.eu/eurostat/statistics-explained/index.php?title=People_in_the_EU_-_population_projections&oldid=497115#Population_projections

[2] A. Meyer, "Contraction and convergence," *Global Commons In*, 2007.
[3] O. Hohmeyer and K. Rennings, *Man-Made Climate Change: Economic Aspects and Policy Options*, vol. 1. Springer Science & Business Media, 2013.
[4] R. Dellink *et al.*, "Sharing the burden of financing adaptation to climate change," *Global Environmental Change*, vol. 19, no. 4, pp. 411–421, 2009, doi: https://doi.org/10.1016/j.gloenvcha.2009.07.009.
[5] N. Hohne, M. Den Elzen, and D. Escalante, "Regional ghg reduction targets based on effort sharing: a comparison of studies," *Climate Policy*, vol. 14, pp. 122–147, 2014.
[6] United Nations, "Paris agreement," 2015, [Online]. Available: http://unfccc.int/paris_agreement/items/9485.php
[7] B. Lahn, "A history of the global carbon budget," *Wiley Interdisciplinary Reviews: Climate Change*, vol. 11, no. 3, p. e636, 2020.
[8] C. Knight, "What is grandfathering?," *Environmental Politics*, vol. 22, no. 3, Art. no. 3, 2013, doi: 10.1080/09644016.2012.740937.
[9] S. Caney, "Justice and the distribution of greenhouse gas emissions," *Journal of Global Ethics*, vol. 5, no. 2, pp. 125–146, 2009, doi: 10.1080/17449620903110300.
[10] S. Caney, "Cosmopolitan Justice, Responsibility, and Global Climate Change," *Leiden Journal of International Law*, vol. 18, no. 4, pp. 747–775, 2005, doi: 10.1017/S0922156505002992.
[11] T. Vadén *et al.*, "Decoupling for ecological sustainability: A categorisation and review of research literature," *Environmental Science & Policy*, vol. 112, pp. 236–244, Oct. 2020, doi: 10.1016/j.envsci.2020.06.016.
[12] European Commission, *Eurostat statistics*. 2022. [Online]. Available: https://ec.europa.eu/eurostat/web/main/home
[13] European Commision, "REGULATION (EU) 2018/842 OF THE EUROPEAN PARLIAMENT AND OF THE COUNCIL." May 30, 2018. [Online]. Available: https://eur-lex.europa.eu/legal-content/EN/TXT/?uri=uriserv:OJ.L_.2018.156.01.0026.01.ENG
[14] European Commission, *'European Climate Law.'* 2021. Accessed: Jan. 30, 2022. [Online]. Available: https://eur-lex.europa.eu/eli/reg/2021/1119/oj
[15] European Energy Agency, "EU Emissions Trading System (ETS) data viewer," Aug. 2021. https://www.eea.europa.eu/data-and-maps/dashboards/emissions-trading-viewer-1
[16] Eurostat, "Greenhouse gas emissions by source sector (source: EEA)," 08 2021. https://ec.europa.eu/eurostat/databrowser/view/env_air_gge/default/table?lang=en
[17] J. Tollefson, "Carbon emissions rapidly rebounded following COVID pandemic dip.," *Nature*, Nov. 2021, doi: 10.1038/d41586-021-03036-x.
[18] Eurostat, "Greenhouse gas emissions in ESD sectors," *Eurostat*, 08 2021. Tables on EU policy > Europe 2020 indicators > Headline indicators > Climate change and energy Cross cutting topics > Climate change > Greenhouse gas emissions
[19] European Commission, *REGULATION (EU) 2018/1999 OF THE EUROPEAN PARLIAMENT AND OF THE COUNCIL*. 2018. Accessed: Feb. 17, 2022. [Online]. Available: https://eur-lex.europa.eu/legal-content/EN/TXT/?toc=OJ:L:2018:328:TOC&uri=uriserv:OJ.L_.2018.328.01.0001.01.ENG
[20] European Commission, "Amending Regulation (EU) 2018/842 on binding annual greenhouse gas emission reductions by Member States from 2021 to 2030 contributing to climate action to meet commitments under the Paris Agreement." 07

2021. [Online]. Available: https://eur-lex.europa.eu/legal-content/EN/TXT/?uri=COM:2021:555:FIN

[21]     European Commission, "Emission Trading System." Jul. 14, 2021. Accessed: Jan. 19, 2022. [Online]. Available: https://ec.europa.eu/clima/eu-action/eu-emissions-trading-system-eu-ets_en

[22]     European Commision, "Amending Directive 2003/87/EC establishing a system for greenhouse gas emission allowance trading within the Union, Decision (EU) 2015/1814 concerning the establishment and operation of a market stability reserve for the Union greenhouse gas emission trading scheme and Regulation (EU) 2015/757." 07 2021. [Online]. Available: https://ec.europa.eu/info/sites/default/files/revision-eu-ets_with_annex_en_0.pdf

[23]     M. S. Kendrick, "The ability-to-pay theory of taxation," *The American Economic Review*, pp. 92–101, 1939.

[24]     S. Sauvé, S. Bernard, and P. Sloan, "Environmental sciences, sustainable development and circular economy: Alternative concepts for trans-disciplinary research," *Environmental Development*, vol. 17, pp. 48–56, 2015, doi: 10.1016/j.envdev.2015.09.002.

[25]     Karl Steininger, Keith Williges, Lukas Meyer, Florian Maczek, and Keywan Riahi, "Sharing the effort of the European Green Deal among countries," *Nature Portfolio*, Mar. 2022, doi: 10.21203/rs.3.rs-1025291/v1.

[26]     Eurostat, "GDP capita at market prices," 04 2022. https://ec.europa.eu/eurostat/databrowser/view/tec00001/default/table?lang=en

[27]     D. Wiedenhofer *et al.*, "A systematic review of the evidence on decoupling of GDP, resource use and GHG emissions, part I: bibliometric and conceptual mapping," *Environmental Research Letters*, vol. 15, no. 6, p. 063002, Jun. 2020, doi: 10.1088/1748-9326/ab8429.

[28]     H. Haberl *et al.*, "A systematic review of the evidence on decoupling of GDP, resource use and GHG emissions, part II: synthesizing the insights," *Environmental Research Letters*, vol. 15, no. 6, p. 065003, Jun. 2020, doi: 10.1088/1748-9326/ab842a.

[29]     P. Tapio, "Towards a theory of decoupling: degrees of decoupling in the EU and the case of road traffic in Finland between 1970 and 2001," *Transport Policy*, vol. 12, no. 2, pp. 137–151, Mar. 2005, doi: 10.1016/j.tranpol.2005.01.001.

[30]     M. R. Raupach *et al.*, "Sharing a quota on cumulative carbon emissions," *Nature Climate Change*, vol. 4, no. 10, Art. no. 10, 2014, doi: 10.1038/nclimate2384.

[31]     S. Wolf, J. Teitge, J. Mielke, F. Schütze, and C. Jaeger, "The European Green Deal — More Than Climate Neutrality," *Intereconomics*, vol. 56, no. 2, pp. 99–107, Mar. 2021, doi: 10.1007/s10272-021-0963-z.

[32]     European Commission, "National long term strategies." Mar. 18, 2022. [Online]. Available: https://ec.europa.eu/info/energy-climate-change-environment/implementation-eu-countries/energy-and-climate-governance-and-reporting/national-long-term-strategies_en#strategies

[33]     V. Duscha, A. Denishchenkova, and J. Wachsmuth, "Achievability of the Paris Agreement targets in the EU: demand-side reduction potentials in a carbon budget perspective," *Climate Policy*, vol. 19, no. 2, pp. 161–174, 2019, doi: 10.1080/14693062.2018.1471385.

[34]     I. Perissi *et al.*, "Potential European Emissions Trajectories within the Global Carbon Budget," pp. 1–13, 2018, doi: 10.3390/su10114225.

[35]   H. D. Matthews, N. P. Gillett, P. A. Stott, and K. Zickfeld, "The proportionality of global warming to cumulative carbon emissions," *Nature*, vol. 459, no. 7248, Art. no. 7248, 2009, doi: 10.1038/nature08047.

[36]   M. R. Allen *et al.*, "Indicate separate contributions of long-lived and short-lived greenhouse gases in emission targets," *npj Climate and Atmospheric Science*, vol. 5, no. 1, p. 5, Jan. 2022, doi: 10.1038/s41612-021-00226-2.

[37]   Our World in Data, "CO2 emissions by region," *Our World in data*, Apr. 14, 2022. https://ourworldindata.org/co2-emissions

[38]   V. Masson-Delmotte, P. Zhai, A. Pirani, and S. L. Connors, "IPCC, 2021: Summary for Policymakers. In: Climate Change 2021: The Physical Science Basis. Contribution of Working Group I to the Sixth Assessment Report of the Intergovernmental Panel on Climate Change." 2021. Accessed: Apr. 23, 2022. [Online]. Available: https://www.ipcc.ch/report/ar6/wg1/downloads/report/IPCC_AR6_WGI_SPM_final.pdf

[39]   Y. Wu, Q. Zhu, and B. Zhu, "Comparisons of decoupling trends of global economic growth and energy consumption between developed and developing countries," *Energy Policy*, vol. 116, pp. 30–38, 2018, doi: https://doi.org/10.1016/j.enpol.2018.01.047.

[40]   I. Perissi, "Highlighting the archetypes of sustainability management by means of simple dynamics models," *Journal of Simulation*, vol. 0, no. 0, Art. no. 0, 2019, doi: 10.1080/17477778.2019.1679612.

[41]   S. Sgouridis, D. Csala, and U. Bardi, "The sower's way: quantifying the narrowing net-energy pathways to a global energy transition," *Environmental Research Letters*, vol. 11, no. 9, Art. no. 9, Sep. 2016, doi: 10.1088/1748-9326/11/9/094009.

[42]   I. Perissi and A. Jones, "Investigating European Union Decarbonization Strategies: Evaluating the Pathway to Carbon Neutrality by 2050," *Sustainability*, vol. 14, no. 8, 2022, doi: 10.3390/su14084728.

[43]   E. Papadis and G. Tsatsaronis, "Challenges in the decarbonization of the energy sector," *Energy*, vol. 205, p. 118025, 2020, doi: https://doi.org/10.1016/j.energy.2020.118025.

[44]   A. Sen, L. Meini, and C. Napoli, *Beyond energy: incentivizing decarbonization through the circular economy*. Oxford Institute for Energy Studies, 2021.

[45]   Eurostat, "Gross domestic product at market prices," 04 2022. https://ec.europa.eu/eurostat/databrowser/view/tec00001/default/table?lang=en

**Appendix A.**

Note: GHG emissions are in MtCO2 eq if not else specified.

**Table A.1.** ESR and ETS historical data for EU 27 as a whole.

| EU 27 - years | ESR historical emissions for EU 27-EUROSTAT (t CO2 eq) | ETS historical emissions for EU 27-EUROSTAT (t CO2 eq) |
|---|---|---|
| 2005 | 2.47E+09 | 1.77E+09 |
| 2006 | 2.46E+09 | 1.78E+09 |
| 2007 | 2.40E+09 | 1.91E+09 |
| 2008 | 2.43E+09 | 1.84E+09 |
| 2009 | 2.34E+09 | 1.63E+09 |
| 2010 | 2.39E+09 | 1.68E+09 |

| 2011 | 2.30E+09 | 1.66E+09 |
|------|----------|----------|
| 2012 | 2.27E+09 | 1.62E+09 |
| 2013 | 2.23E+09 | 1.66E+09 |
| 2014 | 2.15E+09 | 1.59E+09 |
| 2015 | 2.19E+09 | 1.60E+09 |
| 2016 | 2.22E+09 | 1.58E+09 |
| 2017 | 2.25E+09 | 1.59E+09 |
| 2018 | 2.22E+09 | 1.53E+09 |
| 2019 | 2.21E+09 | 1.39E+09 |
| 2020 | 2.08E+09 | 1.22E+09 |

**Table A.2.** ESR emission allocation according to Regulation (EU) 2018/842 of the European Parliament and of the Council (Annex II)

| Regulation-ESR | ESR emissions 2005 | % 2030 | 2021 | 2022 | 2023 | 2024 | 2025 | 2026 | 2027 | 2028 | 2029 | 2030 |
|---|---|---|---|---|---|---|---|---|---|---|---|---|
| Belgium | 8.16E+07 | -35% | 7.11E+07 | 6.91E+07 | 6.71E+07 | 6.51E+07 | 6.31E+07 | 6.11E+07 | 5.91E+07 | 5.71E+07 | 5.51E+07 | 5.30E+07 |
| Bulgaria | 2.23E+07 | 0% | 2.71E+07 | 2.52E+07 | 2.48E+07 | 2.45E+07 | 2.41E+07 | 2.37E+07 | 2.34E+07 | 2.30E+07 | 2.27E+07 | 2.23E+07 |
| Czechia | 6.50E+07 | -14% | 6.60E+07 | 6.09E+07 | 6.03E+07 | 5.97E+07 | 5.90E+07 | 5.84E+07 | 5.78E+07 | 5.71E+07 | 5.65E+07 | 5.59E+07 |
| Denmark | 4.04E+07 | -39% | 3.21E+07 | 3.13E+07 | 3.05E+07 | 2.96E+07 | 2.88E+07 | 2.80E+07 | 2.71E+07 | 2.63E+07 | 2.55E+07 | 2.46E+07 |
| Germany | 4.85E+08 | -38% | 4.27E+08 | 4.13E+08 | 3.99E+08 | 3.85E+08 | 3.71E+08 | 3.57E+08 | 3.43E+08 | 3.29E+08 | 3.15E+08 | 3.01E+08 |
| Estonia | 6.20E+06 | -13% | 6.22E+06 | 6.00E+06 | 5.93E+06 | 5.85E+06 | 5.77E+06 | 5.70E+06 | 5.62E+06 | 5.54E+06 | 5.47E+06 | 5.39E+06 |
| Ireland | 4.77E+07 | -30% | 4.35E+07 | 4.24E+07 | 4.12E+07 | 4.01E+07 | 3.90E+07 | 3.79E+07 | 3.67E+07 | 3.56E+07 | 3.45E+07 | 3.34E+07 |
| Greece | 6.30E+07 | -16% | 4.62E+07 | 4.70E+07 | 4.77E+07 | 4.85E+07 | 4.92E+07 | 4.99E+07 | 5.07E+07 | 5.14E+07 | 5.22E+07 | 5.29E+07 |
| Spain | 2.42E+08 | -26% | 2.01E+08 | 1.99E+08 | 1.96E+08 | 1.94E+08 | 1.92E+08 | 1.89E+08 | 1.87E+08 | 1.85E+08 | 1.82E+08 | 1.80E+08 |
| France | 4.01E+08 | -37% | 3.36E+08 | 3.27E+08 | 3.17E+08 | 3.08E+08 | 2.99E+08 | 2.90E+08 | 2.80E+08 | 2.71E+08 | 2.62E+08 | 2.53E+08 |
| Croatia | 1.81E+07 | -7% | 1.77E+07 | 1.65E+07 | 1.66E+07 | 1.66E+07 | 1.66E+07 | 1.67E+07 | 1.67E+07 | 1.67E+07 | 1.68E+07 | 1.68E+07 |
| Italy | 3.43E+08 | -33% | 2.74E+08 | 2.69E+08 | 2.64E+08 | 2.59E+08 | 2.55E+08 | 2.50E+08 | 2.45E+08 | 2.40E+08 | 2.36E+08 | 2.31E+08 |
| Cyprus | 4.27E+06 | -24% | 4.07E+06 | 3.98E+06 | 3.89E+06 | 3.80E+06 | 3.70E+06 | 3.61E+06 | 3.52E+06 | 3.43E+06 | 3.34E+06 | 3.24E+06 |
| Latvia | 8.60E+06 | -6% | 1.06E+07 | 8.85E+06 | 8.76E+06 | 8.66E+06 | 8.56E+06 | 8.47E+06 | 8.37E+06 | 8.28E+06 | 8.18E+06 | 8.08E+06 |
| Lithuania | 1.31E+07 | -9% | 1.61E+07 | 1.37E+07 | 1.35E+07 | 1.33E+07 | 1.30E+07 | 1.28E+07 | 1.26E+07 | 1.23E+07 | 1.21E+07 | 1.19E+07 |
| Luxembourg | 1.01E+07 | -40% | 8.41E+06 | 8.15E+06 | 7.89E+06 | 7.63E+06 | 7.37E+06 | 7.11E+06 | 6.85E+06 | 6.59E+06 | 6.33E+06 | 6.07E+06 |
| Hungary | 4.78E+07 | -7% | 4.99E+07 | 4.33E+07 | 4.35E+07 | 4.36E+07 | 4.38E+07 | 4.39E+07 | 4.41E+07 | 4.42E+07 | 4.43E+07 | 4.45E+07 |
| Malta | 1.02E+06 | -19% | 2.07E+06 | 1.24E+06 | 1.19E+06 | 1.14E+06 | 1.08E+06 | 1.03E+06 | 9.81E+05 | 9.30E+05 | 8.78E+05 | 8.27E+05 |
| Netherlands | 1.28E+08 | -36% | 9.85E+07 | 9.67E+07 | 9.48E+07 | 9.30E+07 | 9.12E+07 | 8.93E+07 | 8.75E+07 | 8.57E+07 | 8.38E+07 | 8.20E+07 |
| Austria | 5.70E+07 | -36% | 4.88E+07 | 4.74E+07 | 4.60E+07 | 4.47E+07 | 4.33E+07 | 4.19E+07 | 4.06E+07 | 3.92E+07 | 3.78E+07 | 3.65E+07 |
| Poland | 1.92E+08 | -7% | 2.15E+08 | 2.04E+08 | 2.01E+08 | 1.98E+08 | 1.95E+08 | 1.92E+08 | 1.89E+08 | 1.85E+08 | 1.82E+08 | 1.79E+08 |
| Portugal | 4.86E+07 | -17% | 4.25E+07 | 4.08E+07 | 4.08E+07 | 4.07E+07 | 4.07E+07 | 4.06E+07 | 4.06E+07 | 4.05E+07 | 4.05E+07 | 4.04E+07 |
| Romania | 7.82E+07 | -2% | 8.79E+07 | 7.69E+07 | 7.69E+07 | 7.69E+07 | 7.68E+07 | 7.68E+07 | 7.68E+07 | 7.67E+07 | 7.67E+07 | 7.67E+07 |
| Slovenia | 1.18E+07 | -15% | 1.14E+07 | 1.11E+07 | 1.10E+07 | 1.09E+07 | 1.08E+07 | 1.06E+07 | 1.05E+07 | 1.04E+07 | 1.03E+07 | 1.02E+07 |
| Slovakia | 2.31E+07 | -12% | 2.34E+07 | 2.12E+07 | 2.11E+07 | 2.10E+07 | 2.09E+07 | 2.08E+07 | 2.07E+07 | 2.06E+07 | 2.05E+07 | 2.04E+07 |
| Finland | 3.44E+07 | -39% | 2.88E+07 | 2.80E+07 | 2.71E+07 | 2.62E+07 | 2.54E+07 | 2.45E+07 | 2.36E+07 | 2.27E+07 | 2.19E+07 | 2.10E+07 |
| Sweden | 4.32E+07 | -40% | 3.13E+07 | 3.07E+07 | 3.01E+07 | 2.95E+07 | 2.89E+07 | 2.83E+07 | 2.77E+07 | 2.71E+07 | 2.65E+07 | 2.59E+07 |
| Total EU 27 | 2.52E+09 | -30% | 2.23E+09 | 2.14E+09 | 2.10E+09 | 2.06E+09 | 2.01E+09 | 1.97E+09 | 1.93E+09 | 1.88E+09 | 1.84E+09 | 1.80E+09 |

**Table A.3.** Proportioning Member States allocations of Regulation to the objective of Green Deal. For Bulgaria the ratio was not possible and the difference between the emission in 2005 end the emission in 2030 has been divided in 10 equal parts and the quota subtracted yearly from 2021 to 2030 (table 3).

| Member States | % Regulation EC document | % Green Deal EC commission document | Ratio | Emissions 2030 Regulation | Emission 2030 Green Deal |
|---|---|---|---|---|---|
| **Belgium** | -35% | -47% | 1,34 | 8.16E+07 | 4.33E+07 |
| **Bulgaria*** | 0% | -10% | - | 2.23E+07 | 2.01E+07 |
| **Czechia** | -14% | -26% | 1,86 | 6.50E+07 | 4.81E+07 |
| **Denmark** | -39% | -50% | 1,28 | 4.04E+07 | 2.02E+07 |
| **Germany** | -38% | -50% | 1,32 | 4.85E+08 | 2.42E+08 |
| **Estonia** | -13% | -24% | 1,85 | 6.20E+06 | 4.71E+06 |
| **Ireland** | -30% | -42% | 1,40 | 4.77E+07 | 2.77E+07 |
| **Greece** | -16% | -23% | 1,42 | 6.30E+07 | 4.87E+07 |
| **Spain** | -26% | -38% | 1,45 | 2.42E+08 | 1.51E+08 |
| **France** | -37% | -48% | 1,28 | 4.01E+08 | 2.11E+08 |
| **Croatia** | -7% | -17% | 2,39 | 1.81E+07 | 1.50E+07 |
| **Italy** | -33% | -44% | 1,32 | 3.43E+08 | 1.93E+08 |
| **Cyprus** | -24% | -32% | 1,33 | 4.27E+06 | 2.90E+06 |
| **Latvia** | -6% | -17% | 2,83 | 8.60E+06 | 7.14E+06 |
| **Lithuania** | -9% | -21% | 2,33 | 1.31E+07 | 1.03E+07 |
| **Luxembourg** | -40% | -50% | 1,25 | 1.01E+07 | 5.06E+06 |
| **Hungary** | -7% | -19% | 2,67 | 4.78E+07 | 3.89E+07 |
| **Malta** | -19% | -19% | 1,00 | 1.02E+06 | 8.27E+05 |
| **Netherlands** | -36% | -48% | 1,33 | 1.28E+08 | 6.66E+07 |
| **Austria** | -36% | -48% | 1,33 | 5.70E+07 | 2.96E+07 |
| **Poland** | -7% | -18% | 2,53 | 1.92E+08 | 1.58E+08 |
| **Portugal** | -17% | -29% | 1,69 | 4.86E+07 | 3.47E+07 |
| **Romania** | -2% | -13% | 6,35 | 7.82E+07 | 6.83E+07 |
| **Slovenia** | -15% | -27% | 1,80 | 1.18E+07 | 8.63E+06 |
| **Slovakia** | -12% | -23% | 1,89 | 2.31E+07 | 1.79E+07 |
| **Finland** | -39% | -50% | 1,28 | 3.44E+07 | 1.72E+07 |
| **Sweden** | -40% | -50% | 1,25 | 4.32E+07 | 2.16E+07 |
| **total EU 27** | -30% | -40% | 1,33 | 2.52E+09 | 1.51E+09 |

**Table A.4.** Green Deal ESR GHG emission allocations for the Member States between 2021-2030 according to the present study

| Green Deal-ESR Eu Commission | % Emission reduction in 2030 – Green Deal | 2021 | 2022 | 2023 | 2024 | 2025 | 2026 | 2027 | 2028 | 2029 | 2030 |
|---|---|---|---|---|---|---|---|---|---|---|---|
| **Belgium** | -47% | 6,76E+07 | 6,49E+07 | 6,22E+07 | 5,95E+07 | 5,68E+07 | 5,41E+07 | 5,14E+07 | 4,87E+07 | 4,60E+07 | 4,33E+07 |
| **Bulgaria** | -10% | 2,21E+07 | 2,19E+07 | 2,17E+07 | 2,14E+07 | 2,12E+07 | 2,10E+07 | 2,08E+07 | 2,05E+07 | 2,03E+07 | 2,01E+07 |
| **Czechia** | -26% | 6,31E+07 | 5,74E+07 | 5,63E+07 | 5,51E+07 | 5,39E+07 | 5,28E+07 | 5,16E+07 | 5,04E+07 | 4,92E+07 | 4,81E+07 |
| **Denmark** | -50% | 2,98E+07 | 2,87E+07 | 2,77E+07 | 2,66E+07 | 2,55E+07 | 2,45E+07 | 2,34E+07 | 2,23E+07 | 2,13E+07 | 2,02E+07 |
| **Germany** | -50% | 4,09E+08 | 3,91E+08 | 3,72E+08 | 3,54E+08 | 3,35E+08 | 3,16E+08 | 2,98E+08 | 2,79E+08 | 2,61E+08 | 2,42E+08 |
| **Estonia** | -24% | 6,21E+06 | 5,84E+06 | 5,70E+06 | 5,56E+06 | 5,41E+06 | 5,27E+06 | 5,13E+06 | 4,99E+06 | 4,85E+06 | 4,71E+06 |
| **Ireland** | -42% | 4,18E+07 | 4,02E+07 | 3,87E+07 | 3,71E+07 | 3,55E+07 | 3,39E+07 | 3,24E+07 | 3,08E+07 | 2,92E+07 | 2,77E+07 |
| **Greece** | -23% | 3,92E+07 | 4,03E+07 | 4,13E+07 | 4,24E+07 | 4,34E+07 | 4,45E+07 | 4,55E+07 | 4,66E+07 | 4,76E+07 | 4,87E+07 |
| **Spain** | -38% | 1,82E+08 | 1,78E+08 | 1,75E+08 | 1,71E+08 | 1,68E+08 | 1,64E+08 | 1,61E+08 | 1,58E+08 | 1,54E+08 | 1,51E+08 |
| **France** | -48% | 3,17E+08 | 3,05E+08 | 2,93E+08 | 2,82E+08 | 2,70E+08 | 2,58E+08 | 2,46E+08 | 2,34E+08 | 2,22E+08 | 2,11E+08 |
| **Croatia** | -17% | 1,71E+07 | 1,44E+07 | 1,45E+07 | 1,46E+07 | 1,47E+07 | 1,47E+07 | 1,48E+07 | 1,49E+07 | 1,50E+07 | 1,50E+07 |
| **Italy** | -44% | 2,50E+08 | 2,44E+08 | 2,37E+08 | 2,31E+08 | 2,25E+08 | 2,18E+08 | 2,12E+08 | 2,06E+08 | 1,99E+08 | 1,93E+08 |
| **Cyprus** | -32% | 4,01E+06 | 3,89E+06 | 3,76E+06 | 3,64E+06 | 3,52E+06 | 3,39E+06 | 3,27E+06 | 3,15E+06 | 3,02E+06 | 2,90E+06 |

| | | | | | | | | | | |
|---|---|---|---|---|---|---|---|---|---|---|
| Latvia | -17% | 9,32E+06 | 8,69E+06 | 8,65E+06 | 8,62E+06 | 8,50E+06 | 8,23E+06 | 7,96E+06 | 7,68E+06 | 7,41E+06 | 7,14E+06 |
| Lithuania | -21% | 1,44E+07 | 1,33E+07 | 1,32E+07 | 1,30E+07 | 1,30E+07 | 1,25E+07 | 1,19E+07 | 1,14E+07 | 1,09E+07 | 1,03E+07 |
| Luxembourg | -50% | 7,98E+06 | 7,65E+06 | 7,33E+06 | 7,01E+06 | 6,68E+06 | 6,36E+06 | 6,03E+06 | 5,71E+06 | 5,38E+06 | 5,06E+06 |
| Hungary | -19% | 4,86E+07 | 3,58E+07 | 3,62E+07 | 3,66E+07 | 3,70E+07 | 3,74E+07 | 3,77E+07 | 3,81E+07 | 3,85E+07 | 3,89E+07 |
| Malta | -19% | 2,07E+06 | 1,24E+06 | 1,19E+06 | 1,14E+06 | 1,08E+06 | 1,03E+06 | 9,81E+05 | 9,30E+05 | 8,78E+05 | 8,27E+05 |
| Netherlands | -48% | 8,86E+07 | 8,62E+07 | 8,38E+07 | 8,13E+07 | 7,89E+07 | 7,64E+07 | 7,40E+07 | 7,15E+07 | 6,91E+07 | 6,66E+07 |
| Austria | -48% | 4,60E+07 | 4,42E+07 | 4,24E+07 | 4,06E+07 | 3,87E+07 | 3,69E+07 | 3,51E+07 | 3,33E+07 | 3,15E+07 | 2,96E+07 |
| Poland | -18% | 2,01E+08 | 1,97E+08 | 1,96E+08 | 1,95E+08 | 1,93E+08 | 1,90E+08 | 1,82E+08 | 1,74E+08 | 1,66E+08 | 1,58E+08 |
| Portugal | -29% | 3,83E+07 | 3,54E+07 | 3,53E+07 | 3,52E+07 | 3,51E+07 | 3,50E+07 | 3,49E+07 | 3,48E+07 | 3,48E+07 | 3,47E+07 |
| Romania | -13% | 7,98E+07 | 6,98E+07 | 6,97E+07 | 6,95E+07 | 6,93E+07 | 6,91E+07 | 6,89E+07 | 6,87E+07 | 6,85E+07 | 6,83E+07 |
| Slovenia | -27% | 8,76E+06 | 8,73E+06 | 8,72E+06 | 8,71E+06 | 8,69E+06 | 8,68E+06 | 8,67E+06 | 8,66E+06 | 8,65E+06 | 8,63E+06 |
| Slovakia | -23% | 2,33E+07 | 1,94E+07 | 1,92E+07 | 1,90E+07 | 1,88E+07 | 1,86E+07 | 1,84E+07 | 1,83E+07 | 1,81E+07 | 1,79E+07 |
| Finland | -50% | 2,73E+07 | 2,61E+07 | 2,50E+07 | 2,39E+07 | 2,28E+07 | 2,17E+07 | 2,06E+07 | 1,95E+07 | 1,83E+07 | 1,72E+07 |
| Sweden | -50% | 2,84E+07 | 2,76E+07 | 2,69E+07 | 2,61E+07 | 2,54E+07 | 2,46E+07 | 2,39E+07 | 2,31E+07 | 2,24E+07 | 2,16E+07 |
| Total | -40% | 2.08E+09 | 2.21E+09 | 2.01E+09 | 1.95E+09 | 1.89E+09 | 1.82E+09 | 1.76E+09 | 1.70E+09 | 1.64E+09 | 1.58E+09 |

**Table A.5** Normalized GDP capita for EU27 and the Member States with 2010 GDP capita value taken as reference =100.

| | 2010 | 2011 | 2012, | 2013, | 2014 | 2015 | 2016 | 2017 | 2018 | 2019 |
|---|---|---|---|---|---|---|---|---|---|---|
| **European Union 27 countries** | 24.900 | 25.650 | 25.760 | 26.020 | 26.580 | 27.500 | 28.200 | 29.320 | 30.290 | 31.310 |
| | **100,00** | **103,01** | **103,45** | **104,50** | **106,75** | **110,44** | **113,25** | **117,75** | **121,65** | **125,74** |
| **Belgium** | 33.330 | 34.060 | 34.770 | 35.210 | 35.950 | 36.960 | 37.960 | 39.130 | 40.260 | 41.620 |
| | **100,00** | **103,01** | **103,45** | **104,50** | **106,75** | **110,44** | **113,25** | **117,75** | **121,65** | **125,74** |
| **Bulgaria** | 5.080 | 5.640 | 5.780 | 5.790 | 5.960 | 6.380 | 6.840 | 7.420 | 8.000 | 8.820 |
| | **100,00** | **111,02** | **113,78** | **113,98** | **117,32** | **125,59** | **134,65** | **146,06** | **157,48** | **173,62** |
| **Czechia** | 15.020 | 15.740 | 15.470 | 15.170 | 15.000 | 16.080 | 16.790 | 18.330 | 19.850 | 21.140 |
| | **100,00** | **104,79** | **103,00** | **101,00** | **99,87** | **107,06** | **111,78** | **122,04** | **132,16** | **140,75** |
| **Denmark** | 43.840 | 44.500 | 45.530 | 46.100 | 47.090 | 48.050 | 49.420 | 51.140 | 52.180 | 53.370 |
| | **100,00** | **101,51** | **103,85** | **105,16** | **107,41** | **109,60** | **112,73** | **116,65** | **119,02** | **121,74** |
| **Germany** | 31.940 | 33.550 | 34.130 | 34.860 | 36.150 | 37.050 | 38.070 | 39.530 | 40.620 | 41.800 |
| | **100,00** | **105,04** | **106,86** | **109,14** | **113,18** | **116,00** | **119,19** | **123,76** | **127,18** | **130,87** |
| **Estonia** | 11.060 | 12.540 | 13.520 | 14.320 | 15.240 | 15.710 | 16.530 | 18.120 | 19.570 | 20.930 |
| | **100,00** | **113,38** | **122,24** | **129,48** | **137,79** | **142,04** | **149,46** | **163,83** | **176,94** | **189,24** |
| **Ireland** | 36.700 | 37.500 | 38.180 | 38.830 | 41.900 | 55.970 | 56.870 | 61.830 | 67.080 | 72.360 |
| | **100,00** | **102,18** | **104,03** | **105,80** | **114,17** | **152,51** | **154,96** | **168,47** | **182,78** | **197,17** |
| **Greece** | 20.150 | 18.310 | 17.060 | 16.400 | 16.270 | 16.300 | 16.190 | 16.450 | 16.730 | 17.090 |
| | **100,00** | **90,87** | **84,67** | **81,39** | **80,74** | **80,89** | **80,35** | **81,64** | **83,03** | **84,81** |
| **Spain** | 23.040 | 22.760 | 22.050 | 21.900 | 22.220 | 23.220 | 23.980 | 24.970 | 25.750 | 26.420 |
| | **100,00** | **98,78** | **95,70** | **95,05** | **96,44** | **100,78** | **104,08** | **108,38** | **111,76** | **114,67** |
| **France** | 30.690 | 31.510 | 31.820 | 32.080 | 32.420 | 33.020 | 33.430 | 34.230 | 35.070 | 36.050 |
| | **100,00** | **102,67** | **103,68** | **104,53** | **105,64** | **107,59** | **108,93** | **111,53** | **114,27** | **117,46** |
| **Croatia** | 10.610 | 10.600 | 10.420 | 10.420 | 10.370 | 10.740 | 11.320 | 12.080 | 12.880 | 13.660 |
| | **100,00** | **99,91** | **98,21** | **98,21** | **97,74** | **101,23** | **106,69** | **113,85** | **121,39** | **128,75** |
| **Italy** | 26.940 | 27.470 | 26.990 | 26.740 | 26.980 | 27.480 | 28.210 | 28.940 | 29.580 | 30.080 |
| | **100,00** | **101,97** | **100,19** | **99,26** | **100,15** | **102,00** | **104,71** | **107,42** | **109,80** | **111,66** |
| **Cyprus** | 23.400 | 23.270 | 22.500 | 20.880 | 20.450 | 21.100 | 22.230 | 23.550 | 24.840 | 26.090 |
| | **100,00** | **99,44** | **96,15** | **89,23** | **87,39** | **90,17** | **95,00** | **100,64** | **106,15** | **111,50** |
| **Latvia** | 8.550 | 9.550 | 10.870 | 11.320 | 11.850 | 12.430 | 12.950 | 13.900 | 15.130 | 16.020 |
| | **100,00** | **111,70** | **127,13** | **132,40** | **138,60** | **145,38** | **151,46** | **162,57** | **176,96** | **187,37** |
| **Lithuania** | 9.050 | 10.340 | 11.180 | 11.850 | 12.480 | 12.860 | 13.560 | 14.950 | 16.250 | 17.490 |
| | **100,00** | **114,25** | **123,54** | **130,94** | **137,90** | **142,10** | **149,83** | **165,19** | **179,56** | **193,26** |
| **Luxembourg** | 83.550 | 85.330 | 87.540 | 90.030 | 92.760 | 95.090 | 96.230 | 97.440 | 99.150 | 100.890 |
| | **100,00** | **102,13** | **104,78** | **107,76** | **111,02** | **113,81** | **115,18** | **116,62** | **118,67** | **120,75** |
| **Hungary** | 9.980 | 10.250 | 10.110 | 10.340 | 10.770 | 11.460 | 11.850 | 12.980 | 13.920 | 14.950 |
| | **100,00** | **102,71** | **101,30** | **103,61** | **107,92** | **114,83** | **118,74** | **130,06** | **139,48** | **149,80** |
| **Malta** | 16.440 | 16.630 | 17.530 | 18.650 | 20.120 | 22.450 | 23.130 | 25.520 | 26.700 | 27.820 |
| | **100,00** | **101,16** | **106,63** | **113,44** | **122,38** | **136,56** | **140,69** | **155,23** | **162,41** | **169,22** |
| **Netherlands** | 38.470 | 38.960 | 38.970 | 39.300 | 39.820 | 40.730 | 41.590 | 43.090 | 44.920 | 46.880 |
| | **100,00** | **101,27** | **101,30** | **102,16** | **103,51** | **105,87** | **108,11** | **112,01** | **116,77** | **121,86** |
| **Austria** | 35.390 | 36.970 | 37.820 | 38.210 | 38.990 | 39.890 | 40.920 | 42.000 | 43.610 | 44.780 |
| | **100,00** | **104,46** | **106,87** | **107,97** | **110,17** | **112,72** | **115,63** | **118,68** | **123,23** | **126,53** |
| Poland | 9.400 | 9.860 | 10.070 | 10.190 | 10.630 | 11.190 | 11.110 | 12.170 | 12.960 | 13.900 |
| | **100,00** | **104,89** | **107,13** | **108,40** | **113,09** | **119,04** | **118,19** | **129,47** | **137,87** | **147,87** |
| Portugal | 16.990 | 16.680 | 16.010 | 16.300 | 16.640 | 17.350 | 18.060 | 19.020 | 19.950 | 20.840 |
| | **100,00** | **98,18** | **94,23** | **95,94** | **97,94** | **102,12** | **106,30** | **111,95** | **117,42** | **122,66** |
| Romania | 6.200 | 6.540 | 6.620 | 7.190 | 7.570 | 8.080 | 8.630 | 9.580 | 10.500 | 11.520 |
| | **100,00** | **105,48** | **106,77** | **115,97** | **122,10** | **130,32** | **139,19** | **154,52** | **169,35** | **185,81** |
| Slovenia | 17.750 | 18.050 | 17.630 | 17.700 | 18.250 | 18.830 | 19.590 | 20.820 | 22.140 | 23.170 |
| | **100,00** | **101,69** | **99,32** | **99,72** | **102,82** | **106,08** | **110,37** | **117,30** | **124,73** | **130,54** |
| Slovakia | 12.610 | 13.240 | 13.570 | 13.710 | 14.040 | 14.730 | 14.920 | 15.530 | 16.420 | 17.250 |
| | **100,00** | **105,00** | **107,61** | **108,72** | **111,34** | **116,81** | **118,32** | **123,16** | **130,21** | **136,80** |
| Finland | 35.080 | 36.750 | 37.130 | 37.570 | 37.880 | 38.570 | 39.580 | 41.080 | 42.320 | 43.440 |
| | **100,00** | **104,76** | **105,84** | **107,10** | **107,98** | **109,95** | **112,83** | **117,10** | **120,64** | **123,83** |
| Sweden | 39.950 | 43.690 | 45.170 | 46.020 | 45.260 | 46.480 | 46.990 | 47.730 | 46.260 | 46.390 |
| | **100,00** | **109,36** | **113,07** | **115,19** | **113,29** | **116,35** | **117,62** | **119,47** | **115,79** | **116,12** |

**Table A.6** Normalized GHG capita for EU27 and the Member States with 2010 GHG capita value taken as reference =100.

| | 2010 | 2011 | 2012, | 2013, | 2014 | 2015 | 2016 | 2017 | 2018 | 2019 |
|---|---|---|---|---|---|---|---|---|---|---|
| **European Union 27** | 7,1 | 6,6 | 6,3 | 6,0 | 5,8 | 5,7 | 5,7 | 5,5 | 5,4 | 5,2 |
| | 100,00 | 92,96 | 88,73 | 84,51 | 81,69 | 80,28 | 80,28 | 77,46 | 76,06 | 73,24 |
| **Belgium** | 9,7 | 9,5 | 9,3 | 9,1 | 8,7 | 8,8 | 8,8 | 8,9 | 8,7 | 8,4 |
| | 100,00 | 97,94 | 95,88 | 93,81 | 89,69 | 90,72 | 90,72 | 91,75 | 89,69 | 86,60 |
| **Bulgaria** | 12,7 | 11,6 | 11,2 | 11,1 | 10,6 | 10,9 | 10,8 | 10,7 | 10,8 | 10,6 |
| | 100,00 | 91,34 | 88,19 | 87,40 | 83,46 | 85,83 | 85,04 | 84,25 | 85,04 | 83,46 |
| **Czechia** | 8,2 | 8,9 | 8,3 | 7,6 | 8,1 | 8,6 | 8,3 | 8,7 | 8,3 | 8,1 |
| | 100,00 | 108,54 | 101,22 | 92,68 | 98,78 | 104,88 | 101,22 | 106,10 | 101,22 | 98,78 |
| **Denmark** | 13,5 | 13,3 | 12,9 | 12,4 | 12,2 | 12,3 | 12,4 | 12,5 | 12,3 | 11,7 |
| | 100,00 | 98,52 | 95,56 | 91,85 | 90,37 | 91,11 | 91,85 | 92,59 | 91,11 | 86,67 |
| **Germany** | 11,9 | 10,9 | 10,1 | 10,3 | 9,6 | 9,0 | 9,3 | 8,9 | 8,8 | 8,1 |
| | 100,00 | 91,60 | 84,87 | 86,55 | 80,67 | 75,63 | 78,15 | 74,79 | 73,95 | 68,07 |
| **Estonia** | 11,8 | 11,7 | 11,8 | 12,0 | 11,4 | 11,4 | 11,4 | 11,1 | 10,7 | 10,1 |
| | 100,00 | 99,15 | 100,00 | 101,69 | 96,61 | 96,61 | 96,61 | 94,07 | 90,68 | 85,59 |
| **Ireland** | 16,0 | 16,1 | 15,3 | 16,8 | 16,2 | 13,9 | 15,2 | 16,1 | 15,4 | 11,2 |
| | 100,00 | 100,63 | 95,63 | 105,00 | 101,25 | 86,88 | 95,00 | 100,63 | 96,25 | 70,00 |
| **Greece** | 14,1 | 13,1 | 13,2 | 13,1 | 12,9 | 13,4 | 13,7 | 13,6 | 13,5 | 12,8 |
| | 100,00 | 92,91 | 93,62 | 92,91 | 91,49 | 95,04 | 97,16 | 96,45 | 95,74 | 90,78 |
| **Spain** | 10,9 | 10,7 | 10,4 | 9,6 | 9,4 | 9,1 | 8,8 | 9,2 | 9,0 | 8,4 |
| | 100,00 | 98,17 | 95,41 | 88,07 | 86,24 | 83,49 | 80,73 | 84,40 | 82,57 | 77,06 |
| **France** | 8,0 | 7,9 | 7,8 | 7,2 | 7,3 | 7,6 | 7,4 | 7,6 | 7,5 | 7,1 |
| | 100,00 | 98,75 | 97,50 | 90,00 | 91,25 | 95,00 | 92,50 | 95,00 | 93,75 | 88,75 |
| **Croatia** | 8,1 | 7,7 | 7,7 | 7,6 | 7,1 | 7,1 | 7,1 | 7,2 | 6,9 | 6,8 |
| | 100,00 | 95,06 | 95,06 | 93,83 | 87,65 | 87,65 | 87,65 | 88,89 | 85,19 | 83,95 |
| **Italy** | 6,5 | 6,5 | 6,1 | 5,8 | 5,6 | 5,8 | 5,8 | 6,1 | 5,9 | 6,0 |
| | 100,00 | 100,00 | 93,85 | 89,23 | 86,15 | 89,23 | 89,23 | 93,85 | 90,77 | 92,31 |
| **Cyprus** | 8,9 | 8,6 | 8,3 | 7,6 | 7,2 | 7,4 | 7,4 | 7,3 | 7,3 | 7,2 |
| | 100,00 | 96,63 | 93,26 | 85,39 | 80,90 | 83,15 | 83,15 | 82,02 | 82,02 | 80,90 |
| **Latvia** | 12,4 | 11,8 | 11,0 | 10,1 | 10,7 | 10,7 | 11,4 | 11,6 | 11,3 | 11,2 |
| | 100,00 | 95,16 | 88,71 | 81,45 | 86,29 | 86,29 | 91,94 | 93,55 | 91,13 | 90,32 |
| **Lithuania** | 5,8 | 5,5 | 5,5 | 5,5 | 5,5 | 5,6 | 5,7 | 5,8 | 6,1 | 6,1 |
| | 100,00 | 94,83 | 94,83 | 94,83 | 94,83 | 96,55 | 98,28 | 100,00 | 105,17 | 105,17 |
| **Luxembourg** | 6,7 | 7,1 | 7,2 | 6,8 | 6,9 | 7,1 | 7,2 | 7,4 | 7,3 | 7,4 |
| | 100,00 | 105,97 | 107,46 | 101,49 | 102,99 | 105,97 | 107,46 | 110,45 | 108,96 | 110,45 |
| **Hungary** | 26,6 | 25,6 | 24,4 | 22,8 | 21,6 | 20,6 | 20,0 | 20,1 | 20,4 | 20,3 |
| | 100,00 | 96,24 | 91,73 | 85,71 | 81,20 | 77,44 | 75,19 | 75,56 | 76,69 | 76,32 |
| **Malta** | 6,7 | 6,5 | 6,2 | 5,9 | 6,0 | 6,3 | 6,4 | 6,7 | 6,7 | 6,7 |
| | 100,00 | 97,01 | 92,54 | 88,06 | 89,55 | 94,03 | 95,52 | 100,00 | 100,00 | 100,00 |
| **Netherlands** | 7,9 | 7,9 | 8,3 | 7,5 | 7,5 | 5,8 | 5,0 | 5,3 | 5,2 | 5,3 |
| | 100,00 | 100,00 | 105,06 | 94,94 | 94,94 | 73,42 | 63,29 | 67,09 | 65,82 | 67,09 |
| **Austria** | 13,4 | 12,5 | 12,2 | 12,2 | 11,7 | 12,1 | 12,1 | 11,9 | 11,5 | 11,1 |
| | 100,00 | 93,28 | 91,04 | 91,04 | 87,31 | 90,30 | 90,30 | 88,81 | 85,82 | 82,84 |
| **Poland** | 10,3 | 10,0 | 9,7 | 9,6 | 9,2 | 9,3 | 9,4 | 9,6 | 9,2 | 9,3 |
| | 100,00 | 97,09 | 94,17 | 93,20 | 89,32 | 90,29 | 91,26 | 93,20 | 89,32 | 90,29 |
| **Portugal** | 10,9 | 10,9 | 10,7 | 10,6 | 10,3 | 10,3 | 10,6 | 11,0 | 10,9 | 10,4 |
| | 100,00 | 100,00 | 98,17 | 97,25 | 94,50 | 94,50 | 97,25 | 100,92 | 100,00 | 95,41 |
| **Romania** | 6,8 | 6,7 | 6,5 | 6,4 | 6,4 | 6,8 | 6,7 | 7,3 | 6,9 | 6,6 |
| | 100,00 | 98,53 | 95,59 | 94,12 | 94,12 | 100,00 | 98,53 | 107,35 | 101,47 | 97,06 |
| **Slovenia** | 5,9 | 6,3 | 6,2 | 5,8 | 5,8 | 5,9 | 5,8 | 6,0 | 6,1 | 5,9 |
| | 100,00 | 106,78 | 105,08 | 98,31 | 98,31 | 100,00 | 98,31 | 101,69 | 103,39 | 100,00 |
| **Slovakia** | 9,6 | 9,6 | 9,2 | 8,9 | 8,1 | 8,2 | 8,6 | 8,6 | 8,5 | 8,2 |
| | 100,00 | 100,00 | 95,83 | 92,71 | 84,38 | 85,42 | 89,58 | 89,58 | 88,54 | 85,42 |
| **Finland** | 8,4 | 8,3 | 7,9 | 7,8 | 7,4 | 7,5 | 7,6 | 7,8 | 7,8 | 7,4 |
| | 100,00 | 98,81 | 94,05 | 92,86 | 88,10 | 89,29 | 90,48 | 92,86 | 92,86 | 88,10 |
| **Sweden** | 14,4 | 13,0 | 11,9 | 11,9 | 11,1 | 10,4 | 10,9 | 10,4 | 10,7 | 10,1 |

| | 100,00 | 90,28 | 82,64 | 82,64 | 77,08 | 72,22 | 75,69 | 72,22 | 74,31 | 70,14 |

**Table A.7** GDP and GHG capita historical percentage change to to evaluate Tapio Index

| Countries | ΔGDPc 2016-2019 % | ΔGDPc 2017-2019 % | ΔGDPc 2018-2019 % | ΔGHGc 2016-2019 % | ΔGHGc 2017-2019 % | ΔGHGc 2018-2019 % |
|---|---|---|---|---|---|---|
| European Union - 27 | 11.03% | 6.79% | 3,37% | -3,45% | -0,25 | -1,02 |
| Belgium | 9.64% | 6.36% | 3,38% | -1,85% | 0,00 | -0,55 |
| Bulgaria | 28.95% | 18.87% | 10,25% | -2,41% | 0,00 | -0,24 |
| Czechia | 25.91% | 15.33% | 6,50% | -4,88% | -0,13 | -0,75 |
| Denmark | 7.99% | 4.36% | 2,28% | -7,95% | -1,41 | -3,49 |
| Germany | 9.80% | 5.74% | 2,90% | -5,61% | -0,89 | -1,93 |
| Estonia | 26.62% | 15.51% | 6,95% | -27,27% | -0,93 | -3,92 |
| Ireland | 27.24% | 17.03% | 7,87% | -5,19% | -0,03 | -0,66 |
| Greece | 5.56% | 3.89% | 2,15% | -6,67% | -2,19 | -3,10 |
| Spain | 10.18% | 5.81% | 2,60% | -5,33% | -0,20 | -2,05 |
| France | 7.84% | 5.32% | 2,79% | -1,45% | -0,46 | -0,52 |
| Croatia | 20.67% | 13.08% | 6,06% | 1,69% | 0,26 | 0,28 |
| Italy | 6.63% | 3.94% | 1,69% | -1,37% | 0,00 | -0,81 |
| Cyprus | 17.36% | 10.79% | 5,03% | -0,88% | 0,20 | -0,18 |
| Latvia | 23.71% | 15.25% | 5,88% | 0,00% | 0,38 | 0,00 |
| Lithuania | 28.98% | 16.99% | 7,63% | 1,37% | 0,20 | 0,18 |
| Luxembourg | 4.84% | 3.54% | 1,75% | -0,49% | -0,99 | -0,28 |
| Hungary | 26.16% | 15.18% | 7,40% | 0,00% | 0,38 | 0,00 |
| Malta | 20.28% | 9.01% | 4,19% | 1,92% | -1,23 | 0,46 |
| Netherlands | 12.72% | 8.80% | 4,36% | -3,48% | -0,34 | -0,80 |
| Austria | 9.43% | 6.62% | 2,68% | 1,09% | 0,09 | 0,41 |
| Poland | 25.11% | 14.22% | 7,25% | -4,59% | 0,04 | -0,63 |
| Portugal | 15.39% | 9.57% | 4,46% | -4,35% | 0,16 | -0,97 |
| Romania | 33.49% | 20.25% | 9,71% | -3,28% | 0,04 | -0,34 |
| Slovenia | 18.27% | 11.29% | 4,65% | -3,53% | 0,05 | -0,76 |
| Slovakia | 15.62% | 11.08% | 5,05% | -5,13% | 0,00 | -1,01 |
| Finland | 9.75% | 5.74% | 2,65% | -5,61% | -0,71 | -2,12 |
| Sweden | -1.28% | -2.81% | 0,28% | -3,70% | 53,43 | -13,18 |

**Table A.8** Member States Carbon Budget estimation on the base of the effort sharing principle "economic capability.GDP per capita data are last available in the Eurostat series [45] which refers to the year 2020. In Table 2, "GDPcap EU/GDPcap MS" is the ratio between the GDP EU and each member state. The sum of those values, at the bottom of the column GDPcap EU/GDPcap MS represents the summation at the denominator of Eq.1. Then, the capability index is the ratio between GDPcap EU/GDPcap MS and the summation. "CB economic capability" is the result of multiplying the capability index with the total carbon budget of 60 GtCO2eq

| Member States | GDPcap 2020 | GDPcap EU/GDPcap MS | Capability Index | CBs "economic capability" |
|---|---|---|---|---|
| EU27 | 29,910 | | | |
| Belgium | 39,580 | 0.756 | 0.020 | 1226.42 |
| Bulgaria | 8,840 | 3.383 | 0.091 | 5491.15 |
| Czechia | 20,120 | 1.487 | 0.040 | 2412.61 |
| Denmark | 53,600 | 0.558 | 0.015 | 905.63 |
| Germany | 40,490 | 0.739 | 0.020 | 1198.86 |
| Estonia | 20,190 | 1.481 | 0.040 | 2404.25 |
| Ireland | 74,870 | 0.399 | 0.011 | 648.35 |
| Greece | 15,440 | 1.937 | 0.052 | 3143.89 |
| Spain | 23,690 | 1.263 | 0.034 | 2049.04 |
| France | 33,960 | 0.881 | 0.024 | 1429.38 |
| Croatia | 12,400 | 2.412 | 0.065 | 3914.66 |
| Italy | 27,880 | 1.073 | 0.029 | 1741.09 |
| Cyprus | 24,240 | 1.234 | 0.033 | 2002.55 |
| Latvia | 15,480 | 1.932 | 0.052 | 3135.77 |
| Lithuania | 17,710 | 1.689 | 0.046 | 2740.92 |
| Luxembourg | 101,760 | 0.294 | 0.008 | 477.02 |
| Hungary | 14,010 | 2.135 | 0.058 | 3464.79 |
| Malta | 25,320 | 1.181 | 0.032 | 1917.13 |
| Netherlands | 45,870 | 0.652 | 0.018 | 1058.25 |
| Austria | 42,540 | 0.703 | 0.019 | 1141.08 |
| Poland | 13,650 | 2.191 | 0.059 | 3556.17 |
| Portugal | 19,430 | 1.539 | 0.042 | 2498.29 |
| Romania | 11,360 | 2.633 | 0.071 | 4273.04 |
| Slovenia | 22,310 | 1.341 | 0.036 | 2175.78 |
| Slovakia | 16,860 | 1.774 | 0.048 | 2879.11 |
| Finland | 43,030 | 0.695 | 0.019 | 1128.09 |
| Sweden | 45,940 | 0.651 | 0.018 | 1056.63 |
| **Total** | | 37.013 | | 60069.94 |

**Table A.9.** Results on Tapio index estimation, the Decoupling Index, and the related CBs split

| | average Tapio index | rescaled to Malta | $DI_{decj}$ | $1/DI_{decj}$ | | $CB_{MSjdec}$ |
|---|---|---|---|---|---|---|
| **European Union - 27** | -0.75 | -2.01 | | | | |
| **Belgium** | -0.30 | -1.55 | 0.7712 | 1.2966 | 0.0395 | 2262.22 |
| **Bulgaria** | -0.23 | -1.48 | 0.7375 | 1.3560 | 0.0413 | 2572.87 |
| **Czechia** | -0.46 | -1.71 | 0.8541 | 1.1708 | 0.0357 | 2123.53 |
| **Denmark** | -2.39 | -3.64 | 1.8142 | 0.5512 | 0.0168 | 1006.22 |
| **Germany** | -1.55 | -2.81 | 1.3986 | 0.7150 | 0.0218 | 1351.72 |
| **Estonia** | -2.29 | -3.54 | 1.7663 | 0.5662 | 0.0173 | 1031.06 |
| **Ireland** | -0.42 | -1.67 | 0.8308 | 1.2037 | 0.0367 | 2163.37 |
| **Greece** | -2.05 | -3.30 | 1.6459 | 0.6076 | 0.0185 | 1185.22 |
| **Spain** | -1.19 | -2.45 | 1.2189 | 0.8204 | 0.0250 | 1519.57 |
| **France** | -0.70 | -1.95 | 0.9732 | 1.0275 | 0.0313 | 2073.81 |
| **Croatia** | 0.11 | -1.14 | 0.5704 | 1.7531 | 0.0534 | 3425.81 |
| **Italy** | -0.52 | -1.77 | 0.8840 | 1.1312 | 0.0345 | 1990.09 |
| **Cyprus** | -0.20 | -1.45 | 0.7230 | 1.3832 | 0.0422 | 2608.03 |
| **Latvia** | 0.21 | -1.04 | 0.5183 | 1.9294 | 0.0588 | 3215.83 |
| **Lithuania** | 0.09 | -1.16 | 0.5781 | 1.7299 | 0.0527 | 3189.04 |
| **Luxembourg** | 0.10 | -1.15 | 0.5721 | 1.7480 | 0.0533 | 2902.28 |
| **Hungary** | 0.06 | -1.19 | 0.5941 | 1.6833 | 0.0513 | 3069.93 |
| **Malta** | 0.25 | -1.00 | 0.4985 | 2.0061 | 0.0611 | 3952.80 |
| **Netherlands** | -0.74 | -1.99 | 0.9913 | 1.0088 | 0.0307 | 1881.72 |
| **Austria** | -0.06 | -1.31 | 0.6537 | 1.5298 | 0.0466 | 3211.07 |
| **Poland** | -0.36 | -1.62 | 0.8052 | 1.2420 | 0.0378 | 2283.58 |
| **Portugal** | -0.69 | -1.94 | 0.9684 | 1.0326 | 0.0315 | 2066.33 |
| **Romania** | -0.12 | -1.37 | 0.6850 | 1.4598 | 0.0445 | 2600.20 |
| **Slovenia** | -0.48 | -1.73 | 0.8606 | 1.1619 | 0.0354 | 2098.36 |
| **Slovakia** | -0.55 | -1.80 | 0.8973 | 1.1144 | 0.0340 | 2007.83 |
| **Finland** | -1.12 | -2.38 | 1.1844 | 0.8443 | 0.0257 | 1405.25 |
| **Sweden** | -1.46 | -2.71 | 1.3494 | 0.7411 | 0.0226 | 872.20 |
| | | | | 32.8140 | | 60069.94 |

**Table A.10.** EU27 Carbon Budget split by member States based on Inertia principle

| | GHG Emissions 2019 | Inertia Index year 2019 | Carbon Budget due to Inertia term |
|---|---|---|---|
| **European Union - 27** | 3886694 | | |
| **Belgium** | 122330 | 0.0315 | 1890.64 |
| **Bulgaria** | 59013 | 0.0152 | 912.07 |
| **Czechia** | 129660 | 0.0334 | 2003.94 |
| **Denmark** | 50822 | 0.0131 | 785.47 |
| **Germany** | 895454 | 0.2304 | 13839.50 |
| **Estonia** | 19131 | 0.0049 | 295.67 |
| **Ireland** | 64803 | 0.0167 | 1001.56 |
| **Greece** | 94961 | 0.0244 | 1467.66 |
| **Spain** | 345873 | 0.0890 | 5345.56 |
| **France** | 468797 | 0.1206 | 7245.39 |
| **Croatia** | 24471 | 0.0063 | 378.21 |
| **Italy** | 440855 | 0.1134 | 6813.53 |
| **Cyprus** | 9852 | 0.0025 | 152.27 |
| **Latvia** | 11424 | 0.0029 | 176.56 |
| **Lithuania** | 20678 | 0.0053 | 319.59 |
| **Luxembourg** | 12149 | 0.0031 | 187.76 |
| **Hungary** | 64781 | 0.0167 | 1001.22 |
| **Malta** | 2495 | 0.0006 | 38.57 |
| **Netherlands** | 200054 | 0.0515 | 3091.89 |
| **Austria** | 82474 | 0.0212 | 1274.65 |
| **Poland** | 407152 | 0.1048 | 6292.65 |
| **Portugal** | 70901 | 0.0182 | 1095.80 |
| **Romania** | 116327 | 0.0299 | 1797.87 |
| **Slovenia** | 17554 | 0.0045 | 271.30 |
| **Slovakia** | 41585 | 0.0107 | 642.71 |
| **Finland** | 57963 | 0.0149 | 895.84 |
| **Sweden** | 55133 | 0.0142 | 852.10 |

**Table A.11.** Member States Carbon Budgets assignment with a different degree of a blend between Capability and Decoupling for Study 2019.

| | Capability | Capability- Decoupling blended CBs Study 2019 (MtCo2eq) | | | | | | | | | Decoupling |
|---|---|---|---|---|---|---|---|---|---|---|---|
| **Belgium** | 1217 | 1319 | 1421 | 1523 | 1625 | 1727 | 1829 | 1931 | 2033 | 2135 | 2237 |
| **Bulgaria** | 5741 | 5432 | 5123 | 4814 | 4505 | 4196 | 3887 | 3578 | 3269 | 2960 | 2651 |
| **Czechia** | 2395 | 2359 | 2323 | 2286 | 2250 | 2214 | 2178 | 2141 | 2105 | 2069 | 2032 |
| **Denmark** | 949 | 945 | 941 | 936 | 932 | 928 | 924 | 920 | 916 | 912 | 908 |
| **Germany** | 1211 | 1223 | 1234 | 1245 | 1257 | 1268 | 1280 | 1291 | 1302 | 1314 | 1325 |
| **Estonia** | 2419 | 2261 | 2102 | 1944 | 1785 | 1627 | 1468 | 1310 | 1151 | 993 | 834 |
| **Ireland** | 700 | 842 | 984 | 1126 | 1268 | 1410 | 1552 | 1694 | 1836 | 1979 | 2121 |
| **Greece** | 2963 | 2765 | 2567 | 2370 | 2172 | 1974 | 1776 | 1579 | 1381 | 1183 | 985 |
| **Spain** | 1917 | 1853 | 1789 | 1726 | 1662 | 1598 | 1535 | 1471 | 1407 | 1343 | 1280 |
| **France** | 1405 | 1491 | 1578 | 1665 | 1751 | 1838 | 1924 | 2011 | 2098 | 2184 | 2271 |
| **Croatia** | 3707 | 3717 | 3727 | 3738 | 3748 | 3758 | 3769 | 3779 | 3789 | 3799 | 3810 |
| **Italy** | 1683 | 1713 | 1742 | 1772 | 1802 | 1831 | 1861 | 1890 | 1920 | 1949 | 1979 |
| **Cyprus** | 1941 | 2021 | 2102 | 2183 | 2263 | 2344 | 2425 | 2505 | 2586 | 2667 | 2747 |
| **Latvia** | 3161 | 3153 | 3144 | 3136 | 3128 | 3120 | 3111 | 3103 | 3095 | 3087 | 3079 |
| **Lithuania** | 2895 | 2957 | 3018 | 3080 | 3141 | 3203 | 3264 | 3326 | 3388 | 3449 | 3511 |
| **Luxembourg** | 502 | 710 | 918 | 1126 | 1335 | 1543 | 1751 | 1959 | 2167 | 2376 | 2584 |
| **Hungary** | 3387 | 3356 | 3325 | 3294 | 3264 | 3233 | 3202 | 3171 | 3140 | 3109 | 3079 |
| **Malta** | 1820 | 2087 | 2354 | 2621 | 2888 | 3155 | 3422 | 3689 | 3956 | 4223 | 4490 |
| **Netherlands** | 1080 | 1171 | 1262 | 1353 | 1444 | 1535 | 1626 | 1717 | 1808 | 1899 | 1991 |
| **Austria** | 1131 | 1444 | 1757 | 2070 | 2384 | 2697 | 3010 | 3323 | 3636 | 3949 | 4263 |
| **Poland** | 3643 | 3493 | 3344 | 3194 | 3045 | 2895 | 2746 | 2596 | 2446 | 2297 | 2147 |
| **Portugal** | 2430 | 2371 | 2313 | 2254 | 2196 | 2138 | 2079 | 2021 | 1962 | 1904 | 1845 |
| **Romania** | 4396 | 4206 | 4016 | 3827 | 3637 | 3448 | 3258 | 3069 | 2879 | 2690 | 2500 |
| **Slovenia** | 2185 | 2169 | 2153 | 2137 | 2121 | 2105 | 2089 | 2073 | 2057 | 2041 | 2025 |
| **Slovakia** | 2935 | 2823 | 2711 | 2599 | 2487 | 2376 | 2264 | 2152 | 2040 | 1928 | 1816 |
| **Finland** | 1166 | 1175 | 1184 | 1192 | 1201 | 1210 | 1219 | 1228 | 1237 | 1246 | 1255 |
| **Sweden** | 1092 | 1013 | 935 | 856 | 778 | 699 | 621 | 542 | 464 | 385 | 307 |
| **w** | **0** | **0,1** | **0,2** | **0,3** | **0,4** | **0,5** | **0,6** | **0,7** | **0,8** | **0,9** | **1** |

**Table A.12.** Member States Carbon Budgets assignment with a different degree of a blend between Capability and Inertia for Study 2016-2019.

| | Capability | Capability- Inertia blended CBs Study 2016-2019 (MtCo2eq) | | | | | | | | | Inertia |
|---|---|---|---|---|---|---|---|---|---|---|---|
| **Belgium** | 1170 | 1242 | 1314 | 1386 | 1458 | 1530 | 1603 | 1675 | 1747 | 1819 | 1891 |
| **Bulgaria** | 5986 | 5479 | 4971 | 4464 | 3957 | 3449 | 2942 | 2434 | 1927 | 1419 | 912 |
| **Czechia** | 2445 | 2400 | 2356 | 2312 | 2268 | 2224 | 2180 | 2136 | 2092 | 2048 | 2004 |
| **Denmark** | 903 | 891 | 879 | 868 | 856 | 844 | 832 | 821 | 809 | 797 | 785 |
| **Germany** | 1163 | 2430 | 3698 | 4966 | 6233 | 7501 | 8769 | 10036 | 11304 | 12572 | 13840 |
| **Estonia** | 2476 | 2258 | 2040 | 1822 | 1604 | 1386 | 1168 | 950 | 732 | 514 | 296 |
| **Ireland** | 721 | 749 | 777 | 805 | 833 | 861 | 889 | 917 | 945 | 973 | 1002 |
| **Greece** | 2799 | 2666 | 2533 | 2400 | 2267 | 2134 | 2000 | 1867 | 1734 | 1601 | 1468 |
| **Spain** | 1840 | 2190 | 2541 | 2892 | 3242 | 3593 | 3943 | 4294 | 4644 | 4995 | 5346 |
| **France** | 1341 | 1931 | 2522 | 3112 | 3703 | 4293 | 4883 | 5474 | 6064 | 6655 | 7245 |
| **Croatia** | 3725 | 3391 | 3056 | 2721 | 2387 | 2052 | 1717 | 1382 | 1048 | 713 | 378 |
| **Italy** | 1593 | 2115 | 2637 | 3159 | 3681 | 4203 | 4725 | 5247 | 5769 | 6291 | 6814 |
| **Cyprus** | 1924 | 1747 | 1570 | 1392 | 1215 | 1038 | 861 | 684 | 507 | 329 | 152 |
| **Latvia** | 3208 | 2905 | 2602 | 2298 | 1995 | 1692 | 1389 | 1086 | 783 | 480 | 177 |
| **Lithuania** | 2989 | 2722 | 2455 | 2188 | 1921 | 1654 | 1387 | 1120 | 853 | 587 | 320 |
| **Luxembourg** | 473 | 444 | 416 | 387 | 359 | 330 | 302 | 273 | 245 | 216 | 188 |
| **Hungary** | 3465 | 3218 | 2972 | 2726 | 2479 | 2233 | 1987 | 1740 | 1494 | 1248 | 1001 |
| **Malta** | 1803 | 1627 | 1450 | 1274 | 1097 | 921 | 744 | 568 | 392 | 215 | 39 |
| **Netherlands** | 1054 | 1258 | 1462 | 1666 | 1869 | 2073 | 2277 | 2481 | 2684 | 2888 | 3092 |
| **Austria** | 1086 | 1105 | 1124 | 1143 | 1161 | 1180 | 1199 | 1218 | 1237 | 1256 | 1275 |
| **Poland** | 3711 | 3969 | 4227 | 4485 | 4743 | 5002 | 5260 | 5518 | 5776 | 6034 | 6293 |
| **Portugal** | 2389 | 2260 | 2131 | 2001 | 1872 | 1743 | 1613 | 1484 | 1354 | 1225 | 1096 |
| **Romania** | 4625 | 4342 | 4059 | 3777 | 3494 | 3211 | 2929 | 2646 | 2363 | 2081 | 1798 |
| **Slovenia** | 2170 | 1981 | 1791 | 1601 | 1411 | 1221 | 1031 | 841 | 651 | 461 | 271 |
| **Slovakia** | 2902 | 2676 | 2450 | 2224 | 1998 | 1772 | 1546 | 1320 | 1094 | 869 | 643 |
| **Finland** | 1118 | 1096 | 1074 | 1051 | 1029 | 1007 | 985 | 962 | 940 | 918 | 896 |
| **Sweden** | 993 | 979 | 965 | 951 | 937 | 923 | 908 | 894 | 880 | 866 | 852 |
| **w** | **0** | **0,1** | **0,2** | **0,3** | **0,4** | **0,5** | **0,6** | **0,7** | **0,8** | **0,9** | **1** |

**Table A.13.** Member States Carbon Budgets assignment with a different degree of a blend between Capability and Inertia for Study 2019.

| | Capability | Capability- Inertia blended CBs Study 2019 (MtCo2eq) | | | | | | | | | Inertia |
|---|---|---|---|---|---|---|---|---|---|---|---|
| **Belgium** | 1217 | 1291 | 1365 | 1438 | 1512 | 1586 | 1660 | 1734 | 1808 | 1882 | 1956 |
| **Bulgaria** | 5741 | 5258 | 4775 | 4292 | 3809 | 3325 | 2842 | 2359 | 1876 | 1393 | 910 |
| **Czechia** | 2395 | 2356 | 2316 | 2277 | 2237 | 2197 | 2158 | 2118 | 2079 | 2039 | 1999 |
| **Denmark** | 949 | 930 | 911 | 892 | 873 | 855 | 836 | 817 | 798 | 779 | 760 |
| **Germany** | 1211 | 2438 | 3665 | 4891 | 6118 | 7344 | 8571 | 9798 | 11024 | 12251 | 13478 |
| **Estonia** | 2419 | 2201 | 1983 | 1765 | 1547 | 1329 | 1111 | 893 | 675 | 457 | 239 |
| **Ireland** | 700 | 731 | 762 | 794 | 825 | 856 | 888 | 919 | 950 | 982 | 1013 |
| **Greece** | 2963 | 2811 | 2658 | 2506 | 2353 | 2201 | 2049 | 1896 | 1744 | 1591 | 1439 |
| **Spain** | 1917 | 2260 | 2604 | 2948 | 3292 | 3636 | 3980 | 4324 | 4668 | 5012 | 5355 |
| **France** | 1405 | 1994 | 2584 | 3173 | 3763 | 4352 | 4942 | 5532 | 6121 | 6711 | 7300 |
| **Croatia** | 3707 | 3375 | 3043 | 2711 | 2380 | 2048 | 1716 | 1384 | 1052 | 720 | 389 |
| **Italy** | 1683 | 2206 | 2730 | 3253 | 3776 | 4299 | 4822 | 5345 | 5868 | 6391 | 6914 |
| **Cyprus** | 1941 | 1763 | 1584 | 1406 | 1228 | 1050 | 872 | 693 | 515 | 337 | 159 |
| **Latvia** | 3161 | 2863 | 2566 | 2269 | 1971 | 1674 | 1376 | 1079 | 782 | 484 | 187 |
| **Lithuania** | 2895 | 2639 | 2383 | 2126 | 1870 | 1614 | 1358 | 1102 | 845 | 589 | 333 |
| **Luxembourg** | 502 | 472 | 442 | 412 | 382 | 352 | 322 | 292 | 262 | 232 | 202 |
| **Hungary** | 3387 | 3153 | 2919 | 2685 | 2451 | 2217 | 1984 | 1750 | 1516 | 1282 | 1048 |
| **Malta** | 1820 | 1642 | 1465 | 1287 | 1109 | 932 | 754 | 576 | 399 | 221 | 43 |
| **Netherlands** | 1080 | 1281 | 1483 | 1684 | 1885 | 2087 | 2288 | 2489 | 2691 | 2892 | 3093 |
| **Austria** | 1131 | 1151 | 1170 | 1190 | 1210 | 1230 | 1249 | 1269 | 1289 | 1309 | 1329 |
| **Poland** | 3643 | 3911 | 4179 | 4447 | 4715 | 4983 | 5251 | 5519 | 5787 | 6055 | 6323 |
| **Portugal** | 2430 | 2296 | 2162 | 2028 | 1895 | 1761 | 1627 | 1493 | 1360 | 1226 | 1092 |
| **Romania** | 4396 | 4139 | 3883 | 3627 | 3371 | 3115 | 2859 | 2603 | 2347 | 2091 | 1835 |
| **Slovenia** | 2185 | 1994 | 1803 | 1612 | 1421 | 1230 | 1039 | 848 | 657 | 466 | 275 |
| **Slovakia** | 2935 | 2706 | 2477 | 2248 | 2019 | 1790 | 1561 | 1332 | 1103 | 874 | 645 |

| | | | | | | | | | | |
|---|---|---|---|---|---|---|---|---|---|---|
| **Finland** | 1166 | 1138 | 1111 | 1084 | 1057 | 1030 | 1002 | 975 | 948 | 921 | 893 |
| **Sweden** | 1092 | 1068 | 1045 | 1022 | 999 | 976 | 953 | 930 | 907 | 883 | 860 |
| **w** | **0** | **0,1** | **0,2** | **0,3** | **0,4** | **0,5** | **0,6** | **0,7** | **0,8** | **0,9** | **1** |

**Table A.14.** Member States Carbon Budgets assignment with a different degree of a blend between Capability, Decoupling, and Inertia for Study 2016-2019.

| | Capability | Capability- Decoupling-Inertia blended CBs Study 2016-2019 (MtCo2eq) | | | | | | | | | Inertia |
|---|---|---|---|---|---|---|---|---|---|---|---|
| **Belgium** | 1170 | 1242 | 1314 | 1386 | 1458 | 1530 | 1603 | 1675 | 1747 | 1819 | 1891 |
| **Bulgaria** | 5986 | 5479 | 4971 | 4464 | 3957 | 3449 | 2942 | 2434 | 1927 | 1419 | 912 |
| **Czechia** | 2445 | 2400 | 2356 | 2312 | 2268 | 2224 | 2180 | 2136 | 2092 | 2048 | 2004 |
| **Denmark** | 903 | 891 | 879 | 868 | 856 | 844 | 832 | 821 | 809 | 797 | 785 |
| **Germany** | 1163 | 2430 | 3698 | 4966 | 6233 | 7501 | 8769 | 10036 | 11304 | 12572 | 13840 |
| **Estonia** | 2476 | 2258 | 2040 | 1822 | 1604 | 1386 | 1168 | 950 | 732 | 514 | 296 |
| **Ireland** | 721 | 749 | 777 | 805 | 833 | 861 | 889 | 917 | 945 | 973 | 1002 |
| **Greece** | 2799 | 2666 | 2533 | 2400 | 2267 | 2134 | 2000 | 1867 | 1734 | 1601 | 1468 |
| **Spain** | 1840 | 2190 | 2541 | 2892 | 3242 | 3593 | 3943 | 4294 | 4644 | 4995 | 5346 |
| **France** | 1341 | 1931 | 2522 | 3112 | 3703 | 4293 | 4883 | 5474 | 6064 | 6655 | 7245 |
| **Croatia** | 3725 | 3391 | 3056 | 2721 | 2387 | 2052 | 1717 | 1382 | 1048 | 713 | 378 |
| **Italy** | 1593 | 2115 | 2637 | 3159 | 3681 | 4203 | 4725 | 5247 | 5769 | 6291 | 6814 |
| **Cyprus** | 1924 | 1747 | 1570 | 1392 | 1215 | 1038 | 861 | 684 | 507 | 329 | 152 |
| **Latvia** | 3208 | 2905 | 2602 | 2298 | 1995 | 1692 | 1389 | 1086 | 783 | 480 | 177 |
| **Lithuania** | 2989 | 2722 | 2455 | 2188 | 1921 | 1654 | 1387 | 1120 | 853 | 587 | 320 |
| **Luxembourg** | 473 | 444 | 416 | 387 | 359 | 330 | 302 | 273 | 245 | 216 | 188 |
| **Hungary** | 3465 | 3218 | 2972 | 2726 | 2479 | 2233 | 1987 | 1740 | 1494 | 1248 | 1001 |
| **Malta** | 1803 | 1627 | 1450 | 1274 | 1097 | 921 | 744 | 568 | 392 | 215 | 39 |
| **Netherlands** | 1054 | 1258 | 1462 | 1666 | 1869 | 2073 | 2277 | 2481 | 2684 | 2888 | 3092 |
| **Austria** | 1086 | 1105 | 1124 | 1143 | 1161 | 1180 | 1199 | 1218 | 1237 | 1256 | 1275 |
| **Poland** | 3711 | 3969 | 4227 | 4485 | 4743 | 5002 | 5260 | 5518 | 5776 | 6034 | 6293 |
| **Portugal** | 2389 | 2260 | 2131 | 2001 | 1872 | 1743 | 1613 | 1484 | 1354 | 1225 | 1096 |
| **Romania** | 4625 | 4342 | 4059 | 3777 | 3494 | 3211 | 2929 | 2646 | 2363 | 2081 | 1798 |
| **Slovenia** | 2170 | 1981 | 1791 | 1601 | 1411 | 1221 | 1031 | 841 | 651 | 461 | 271 |
| **Slovakia** | 2902 | 2676 | 2450 | 2224 | 1998 | 1772 | 1546 | 1320 | 1094 | 869 | 643 |
| **Finland** | 1118 | 1096 | 1074 | 1051 | 1029 | 1007 | 985 | 962 | 940 | 918 | 896 |
| **Sweden** | 993 | 979 | 965 | 951 | 937 | 923 | 908 | 894 | 880 | 866 | 852 |
| **w** | **0** | **0,1** | **0,2** | **0,3** | **0,4** | **0,5** | **0,6** | **0,7** | **0,8** | **0,9** | **1** |

**Table A.15.** Member States Carbon Budgets assignment with a different degree of a blend between Capability, Decoupling, and Inertia for Study 2019.

| | Capability | Capability- Decoupling-Inertia blended CBs Study 2019 (MtCo2eq) | | | | | | | | | Inertia |
|---|---|---|---|---|---|---|---|---|---|---|---|
| **Belgium** | 1217 | 1382 | 1528 | 1653 | 1757 | 1842 | 1905 | 1949 | 1971 | 1974 | 1956 |
| **Bulgaria** | 5741 | 4980 | 4280 | 3643 | 3067 | 2553 | 2101 | 1710 | 1382 | 1115 | 910 |
| **Czechia** | 2395 | 2323 | 2258 | 2200 | 2150 | 2107 | 2071 | 2042 | 2021 | 2006 | 1999 |
| **Denmark** | 949 | 926 | 905 | 884 | 864 | 844 | 826 | 808 | 791 | 775 | 760 |
| **Germany** | 1211 | 2448 | 3683 | 4915 | 6145 | 7373 | 8598 | 9822 | 11042 | 12261 | 13478 |
| **Estonia** | 2419 | 2059 | 1730 | 1432 | 1167 | 933 | 731 | 560 | 422 | 315 | 239 |
| **Ireland** | 700 | 859 | 990 | 1092 | 1166 | 1212 | 1229 | 1218 | 1178 | 1110 | 1013 |
| **Greece** | 2963 | 2633 | 2342 | 2090 | 1879 | 1707 | 1574 | 1481 | 1427 | 1413 | 1439 |
| **Spain** | 1917 | 2203 | 2502 | 2815 | 3139 | 3477 | 3827 | 4190 | 4566 | 4954 | 5355 |
| **France** | 1405 | 2072 | 2722 | 3355 | 3971 | 4569 | 5150 | 5714 | 6260 | 6789 | 7300 |
| **Croatia** | 3707 | 3384 | 3060 | 2733 | 2404 | 2073 | 1741 | 1406 | 1069 | 730 | 389 |
| **Italy** | 1683 | 2233 | 2777 | 3315 | 3847 | 4373 | 4893 | 5407 | 5915 | 6418 | 6914 |
| **Cyprus** | 1941 | 1835 | 1713 | 1576 | 1422 | 1251 | 1065 | 863 | 644 | 409 | 159 |
| **Latvia** | 3161 | 2856 | 2553 | 2251 | 1951 | 1653 | 1357 | 1062 | 768 | 477 | 187 |
| **Lithuania** | 2895 | 2694 | 2481 | 2256 | 2018 | 1768 | 1506 | 1231 | 944 | 645 | 333 |
| **Luxembourg** | 502 | 659 | 775 | 849 | 881 | 872 | 821 | 729 | 595 | 419 | 202 |
| **Hungary** | 3387 | 3125 | 2870 | 2621 | 2377 | 2140 | 1910 | 1685 | 1466 | 1254 | 1048 |
| **Malta** | 1820 | 1883 | 1892 | 1848 | 1750 | 1599 | 1395 | 1137 | 826 | 461 | 43 |

| | | | | | | | | | | |
|---|---|---|---|---|---|---|---|---|---|---|
| **Netherlands** | 1080 | 1363 | 1628 | 1875 | 2104 | 2314 | 2507 | 2681 | 2836 | 2974 | 3093 |
| **Austria** | 1131 | 1432 | 1671 | 1848 | 1962 | 2013 | 2001 | 1927 | 1790 | 1591 | 1329 |
| **Poland** | 3643 | 3776 | 3940 | 4133 | 4356 | 4609 | 4892 | 5205 | 5548 | 5921 | 6323 |
| **Portugal** | 2430 | 2243 | 2069 | 1906 | 1754 | 1615 | 1487 | 1371 | 1266 | 1173 | 1092 |
| **Romania** | 4396 | 3969 | 3580 | 3229 | 2916 | 2641 | 2404 | 2205 | 2044 | 1920 | 1835 |
| **Slovenia** | 2185 | 1980 | 1778 | 1579 | 1383 | 1190 | 1001 | 815 | 632 | 452 | 275 |
| **Slovakia** | 2935 | 2606 | 2298 | 2013 | 1750 | 1510 | 1292 | 1097 | 924 | 773 | 645 |
| **Finland** | 1166 | 1146 | 1126 | 1103 | 1078 | 1052 | 1024 | 994 | 962 | 929 | 893 |
| **Sweden** | 1092 | 998 | 920 | 857 | 811 | 780 | 764 | 765 | 781 | 813 | 860 |
| **w** | **0** | **0,1** | **0,2** | **0,3** | **0,4** | **0,5** | **0,6** | **0,7** | **0,8** | **0,9** | **1** |